\documentclass[english,floats,twocolumn]{revtex4-2}
\usepackage[T1]{fontenc}
\usepackage[utf8]{inputenc}
\usepackage{color}
\usepackage{babel}
\usepackage{amssymb}
\usepackage{graphicx}
\usepackage[pdfusetitle,
 bookmarks=true,bookmarksnumbered=false,bookmarksopen=false,
 breaklinks=false,pdfborder={0 0 0},pdfborderstyle={},backref=false,colorlinks=true]
 {hyperref}

\makeatletter

\usepackage{babel}
\usepackage{inputenc}

\makeatother

\begin{document}
\title{Energy minima and ordering in ferromagnets with quenched randomness}
\author{D. A. Garanin}
\affiliation{Physics Department, Herbert H. Lehman College and Graduate School,
The City University of New York, 250 Bedford Park Boulevard West,
Bronx, New York 10468-1589, USA}
\date{\today}
\begin{abstract}
Energy minimization at $T=0$ and Monte Carlo simulations at $T>0$
have been performed for 2D and 3D random-field and random-anisotropy
systems of up to 100 million classical spins. The main finding is
that 3D random-anisotropy systems magnetically order on lowering temperature,
contrary to the theoretical predictions based on the Imry-Ma argument.
If random-anisotropy is stronger than the exchange, which can be the
case in sintered materials, the system still orders but the magnetization
is strongly reduced and there is a large spin-glass component in the
spin state, the heat capacity having a cusp instead of a divergence.
3D random-field systems do not magnetically order on lowering temperature
but rather freeze into the correlated spin-glass state. Here, although
magnetized local energy minima have lower energies than non-magnetized
ones, magnetic ordering is prevented by singularities pinned by the
random field.
\end{abstract}
\maketitle

\section{Introduction}

Quenched or static randomness is ubiquitous in solid-state physics,
strongly affecting different types of ordering. In magnetic systems,
randomness arises via lattice defects as well as random anisotropy,
which is usually due to the amorphous or sintered structure of these
materials \citep{marher00jmmm,bilcantam05prb}. Other physical systems
include flux lattices in superconductors \citep{lar70jetp,chu89prb,blafeigeslarvin94rmp},
thin films on imperfect substrates \citep{chu86prb}, pinned charge-density
waves \citep{efelar77jetp}, superfluid $^{3}$He-A in aerogel \citep{vol08jltp,lipolzimcolganhal13nature},
magnetic-bubble lattices \citep{seswes92prb_1,seswes92prb_2}, liquid
crystals \citep{belcladegmannat98pre} to name a few. Larkin, with
the help of the Green's functions (GF) method, had shown that whatever
weak pinning potentials destroy the long-range translational order
of flux lattices \citep{lar70jetp}. The same method was applied to
charge-density waves \citep{efelar77jetp}.

Important classes of systems with quenched randomness are systems
with random field (RF) and random anisotropy (RA). For magnetic systems,
the most relevant is the RA model as the crystal field produces only
even combinations of the spin-vector components due to the time-reversal
symmetry. In theoretical investigations, one usually uses the simplified
model of spins on a regular lattice with the RA \citep{harplizuc73prl}.
An effective random field arises in diluted antiferromagnets in the
presence of an applied uniform magnetic field \citep{fisaha79jpcm}.
A random field can be used to model pinning centers in different systems.
Apart from that, the RF model is important theoretically as the most
basic model of static randomness in magnetic systems.

The interest in systems with quenched randomness increased after the
seminal work by Imry and Ma \citep{imrma75prl}, in which suppression
of long-range order (LRO) in magnetic systems with continuous symmetry
by whatever weak random field was suggested based on a simple energy
argument. This amounts to the absence of magnetization in the ground
state of these systems. Aizenman and Wehr \citep{aizweh89prl,aizweh90cmp}
provided a mathematical argument that is considered to be a rigorous
proof of the Imry-Ma statement. Imry-Ma's simple approach leads to
qualitatively the same results as the sophisticated GF method by Larkin.
Both yield a correlation length that is finite at $T=0$ and diverges
in the limit of weak randomness, signaling the absence of the LRO.
This is why both together are sometimes called the Larkin-Imry-Ma
(LIM) argument, albeit theoretically they have little in common. So
below the abbreviation IM for the Imry-Ma argument will be used.

Random-anisotropy model was investigated with the help of the mean-field
theory \citep{calliucul77prb}, Monte Carlo \citep{jaykir80prb},
perturbation theory \citep{ahapyt80}, and scaling \citep{ahapyt83}.
Early work was summarized in Ref. \citep{aha83jmmm}, as well as in
a much later Ref. \citep{dudfolhol05jmmm}, which uses the renormalization-group
method. In particular, Aharony and Pytte \citep{ahapyt80}, inspired
by the Imry-Ma argument, stated that the RA model has zero magnetization
and a huge magnetic susceptibility at low temperatures.

The GF method was applied to the classical 3D $xy$ ferromagnetic
RF model in Ref. \citep{garchupro13prb} and to the 3D arbitrary-component
spin-vector RF model in Refs. \citep{progarchu14prl,garchu15epjb},
in both cases with parallel direct numerical energy minimization for
spins on the lattice. This research to a significant extent was fueled
by the idea to numerically check the theoretical work criticizing
the Imry-Ma argument \citep{carost82prb,vilfer84zfb} and proposing
a power-law dependence of the correlation function based on the renormalization
group with replicas \citep{kor93prb,giadou94prl,giadou95prb,fel00prb,fel01ijmp,dou06prl}
(for a review, see Ref. \citep{natsch00advphys}), variational approach
\citep{gariororl96prb}, as well as Monte Carlo simulations \citep{ginhus96prb,fis98prb,fis00prb,ita03prb}.
In particular, Refs. \citep{jaykir80prb,ginhus96prb,fel00prb,fel01ijmp,gariororl96prb,fis98prb,fis00prb,ita03prb}
deal with spin systems. This hypothetical quasiordered phase, presumed
to be vortex-free in spin systems and dislocation-free in flux lattices,
received the name of a Bragg glass (see, for instance, Ref. \citep{dou06prl}).
In Refs. \citep{garchupro13prb,garchu15epjb} it was found that at
short distances the spin correlation functions (CF) follow the predictions
of the GF theory, and the numerically found values of the ferromagnetic
correlation radius $R_{f}$ coincide with those obtained by the GF
method. No power-law dependences were observed at large distances,
$R\gtrsim R_{f}$.

It was suggested, although never proven rigorously or demonstrated
numerically on large systems, that the RA model behaves similarly
to the RF model. The GF method was applied to RA ferromagnets in Refs.
\citep{chuser82prb,chusasser86prb,chu86prb_ra} much earlier than
it was applied to RF ferromagnets. Similarly to the latter and Ref.
\citep{ahapyt80}, it was conjectured that there is no LRO in RA ferromagnets
and $R_{f}$ is large for a weak RA, which implies an extreme softness
(a large magnetic susceptibility) of amorphous and sintered magnets.
Such a state was called a \textit{correlated spin glass.} Magnetic
hysteresis in the 3D RA ferromagnet, which is not captured by the
GF method, was studied by numerical energy minimization in Ref. \citep{prochugar15jmmm_RA}.
Freezing of the 2D RA ferromagnet into a correlated spin glass was
investigated in Ref. \citep{garchu22jpcm} by Monte Carlo simulations,
and it was inferred that the freezing mechanism is likely blocking
of spin flips by anisotropy barriers rather than a true spin-glass
transition.

Possible applications of RA magnets include very soft magnetic elements,
as predicted in Refs. \citep{chuser82prb,chusasser86prb,chu86prb_ra},
and efficient broadband absorbers of the microwave energy \citep{garchu21prb,garchu22prb,garchu23prb_IP}.
Although at the atomic scale anisotropy relative to the exchange is
small as a relativistic effect, it is largely augmented in granular
materials where directions of the random-anisotropy axes are the same
within grains \citep{prochugar15jmmm_RA,chugar24prb_sc}.

Numerical results of Refs. \citep{garchupro13prb,prochugar15jmmm_RA,garchu15epjb}
uncovered another aspect of the behavior of systems with quenched
randomness. Both RF and RA models possess zillions of local energy
minima separated by energy barriers. As a result of energy minimization
or processes on the system (lowering the temperature, sweeping the
magnetic field, etc.) the system can end up in any of these states,
not necessarily in the ground state. Energy minimization of the RF
model starting from the random initial condition (RIC) for spins leads
to the spin-glass state with nearly zero magnetization. The same starting
from the collinear initial condition (CIC) leads to a partially ordered
state. In this case, the spin CF decreases with $R_{f}$ calculated
by the GF method, but then flattens out, revealing incomplete disordering
with a significant residual magnetization. If the amplitude of the
RF is small enough, this state is free of singularities (vortices,
vortex loops, hedgehogs), which are ignored in the GF approach, as
well as in the Imry-Ma argument. In the case of RIC, there are lots
of singularities that increase the energy of the system and make the
magnetization vanish. Magnetic correlation radius in this case (called
$R_{V}$ in Ref. \citep{garchupro13prb} because of vortex loops in
the 3D $xy$ model) is controlled by the concentration of singularities
and is significantly smaller than $R_{f}$ resulting from the GF calculation.

Singularities play a major role in the ordering/disordering of systems
with quenched randomness. In the last section of Ref. \citep{garchupro13prb},
and then in a more general and focused form in Ref. \citep{progarchu14prl},
if was shown that, if the spins follow the random field averaged over
the region of the linear size $R_{f}$, as assumed in the Imry-Ma
argument, this inevitably leads to formation of singularities: vortices
in the 2D $xy$ model, vortex loops in the 3D $xy$ model, hedgehogs
in the 3D Heisenberg model. In the 2D Heisenberg model, there are
no singularities, but there are nonsingular topological objects --
skyrmions. For the general $n$-component spins in $d$ dimensions,
at $n\le d$ the presence of singularities leads to strong metastability
and glassy behavior, with the final state depending on the initial
condition. At $n=d+1$, when topological structures are nonsingular,
the system possesses a weak metastability. At $n>d+1$, when topological
objects are absent, the final, lowest-energy state is independent
of the initial condition. It is characterized by the exponential decay
of correlations that agrees quantitatively with the theory based upon
the Imry-Ma argument. One can generalize these findings, stating that
any true disordering (excluding quasi-regular structures with zero
magnetization, such as stripe domains) leads to the formation of singularities.
The role of the latter may be preventing complete disordering of the
system if the increase of the energy due to singularities dominates
the energy decrease due to further aligning with the RF. Results of
Refs. \citep{garchupro13prb,progarchu14prl,garchu15epjb} indicate
that all RF systems disorder except for the 3D $xy$ model, in which
the energy of creating vortex loops is very high.

So far, numerical investigations of the RA model are scarce in comparison
to those of the RF model. One can name Refs. \citep{prochugar15jmmm_RA,garchu22jpcm}
cited above, as well as the Monte Carlo study of a spin chain with
RA \citep{dicchu91prb} and small systems with RF and RA in Refs.
\citep{fis98prb,fis00prb}, despite the RA systems being more relevant
for applications. This can be partially explained by the belief that
RA has a similar effect to RF. However, there is an important difference
-- the degeneracy of the RA energy with respect to the inversion
of spins. Whereas in the RF model energetically favorable directions
of spins cover the whole sphere, in the RA model the directions of
the spins can be reduced to a hemisphere, which minimizes the exchange
energy. This is the case of the strong RA, in which the spin polarization
in the ground state is $m=1/2$ for unit-length classical spin vectors.
This differs radically from the predictions based on the Imry-Ma argument
and the GF approach made for a weak RA. However, weakening RA relative
to the exchange interaction is unlikely to increase disordering --
one can expect the opposite. Based on the energy argument above, one
can expect that RA systems will order with lowering temperature, although
the equilibration can be slowed down by energy barriers, especially
at low $T$.

The objective of this work is to clarify the question of ordering
in magnetic systems with static disorder, especially in RA systems,
with the help of energy minimization at $T=0$ and Monte Carlo simulations
at finite temperatures. This requires working with large systems as
the magnetic correlation radius $R_{f}$ can be large, and the system
size $L$ should satisfy $L\gg R_{f}$ to avoid ordering as a trivial
finite-size effect. Current computing resources make it possible.

The structure of the remaining part of the paper is the following.
In Sec. \ref{Sec_The-model} the model of classical ferromagnets with
RF and RA is introduced, and the theoretical background based on the
Imry-Ma argument and Green's function approach is reviewed. In Sec.
\ref{Sec_Numerical-procedures} the numerical methods, the energy
minimization, and the Monte Carlo with thermalized overrelaxation
are explained. Section \ref{Sec_Numerical-results} contains the numerical
results subdivided into three groups: (i) Sampled local energy minima
in RA and RF systems in 2D and 3D; (ii) $D_{R}$ and $H_{R}$ scans
at $T=0$ which reveal the change of the ground state; (iii) temperature
dependences of the magnetization and other physical quantities. The
results are summarized in the Conclusion.

\section{The model}

\label{Sec_The-model}

Consider the model of classical spins on a square or cubic lattice

\begin{eqnarray}
\mathcal{H} & = & -\frac{1}{2}\sum_{i,j}J_{ij}\mathbf{s}_{i}\cdot{\bf s}_{j}-H_{R}\sum_{i}({\bf n}_{i}\cdot{\bf s}_{i})\nonumber \\
 &  & -\frac{D_{R}}{2}\sum_{i}({\bf n}_{i}\cdot{\bf s}_{i})^{2}-\mathbf{H}\cdot\sum_{i}{\bf s}_{i}\label{Hamiltonian}
\end{eqnarray}
with either RF (second term) or RA (third term) taken into account.
Here $J_{ij}$ is the coupling of the $n$-component spin vectors
$\left|{\bf s}_{i}\right|=1$ with the coupling constant $J>0$ between
the nearest neighbors, $H_{R}>0$ is the RF constant, $D_{R}>0$ is
the RA constant, ${\bf n}_{i}$ are randomly oriented unit vectors,
and $\mathbf{H}$ is the applied field in the energy units.

Before performing computations on RF and RA systems, one needs to
know the magnetic correlation radius $R_{f}$ to ensure it is smaller
than the system's size $L$. The easiest way to estimate $R_{f}$
is using the Imry-Ma argument. The latter in the RF case assumes that
spins within a region of up to the linear size $R_{f}$ are aligned
with the random field averaged over this region that can be estimated
as $H_{R}\left(a/R_{f}\right)^{d}$, where $a$ is the lattice spacing
and $d$ is the dimensionality. The direction of the spins changes
at the characteristic distance $R_{f}$, which costs the exchange
energy via the gradient of the spin field that can be estimated as
$J\left(a/R_{f}\right)^{2}$ per spin. Both RF and exchange energies
together are given by
\begin{equation}
E=-H_{R}\left(\frac{a}{R_{f}}\right)^{d/2}+J\left(\frac{a}{R_{f}}\right)^{2}.
\end{equation}
where the coefficients in both terms are undefined. Minimizing $E$
with respect to $R_{f}$ yields
\begin{equation}
\frac{R_{f}}{a}=k_{d}\left(\frac{J}{H_{R}}\right)^{2/(4-d)},\qquad E=-k_{d}'\frac{H_{R}^{4/(4-d)}}{J^{d/(4-d)}},\label{Rf}
\end{equation}
where $k_{d}$ and $k_{d}'$ are numerical factors. The result is
that a ferromagnet with RF contains strongly correlated regions, which
can be called \textit{Imry-Ma domains}, although there are no boundaries
between them, and thus they can be chosen in different ways. The ferromagnetic
correlation length $R_{f}$ is finite for all $d<4$, no matter how
weak the RA is. This is usually interpreted as proof of the absence
of ordering in the RF system. However, a finite $R_{f}$ can exist
in a partially ordered system, for instance, in the states obtained
by the energy minimization from CIC (see the CF figures in Refs. \citep{garchupro13prb,garchu15epjb}).
On the other hand, the absence of ordering is assumed in the IM construction
(spins following the averaged RF). The GF calculations yield qualitatively
the same results as the Imry-Ma argument. In 3D one obtains \citep{progarchu14prl,garchu15epjb}
\begin{equation}
\frac{R_{f}}{a}=\frac{8\pi}{1-1/n}\left(\frac{J}{H_{R}}\right)^{2}\label{Rf_RF}
\end{equation}
and \citep{garchu15epjb}
\begin{equation}
E=-\frac{H_{R}^{4}}{128\pi^{2}J^{3}},\qquad n\rightarrow\infty.\label{E_IM}
\end{equation}
The coefficient $k_{d}$ following from Eq. (\ref{Rf_RF}) is large
but one obtains a coefficient of order one if one measures $H_{R}$
against $6J$, which is the total exchange field from all 6 neighbors.
The very small Imry-Ma adjustment energy $E$ could be calculated
analytically only for the mean spherical model, $n\rightarrow\infty$.
In fact, the spin field is not smooth, as assumed in the IM construction,
but contains a short-range component adjusting to the RF at the atomic
scale $a$. This short-range energy in 3D is given by \citep{garchu15epjb}
\begin{equation}
E_{\mathrm{SR}}=-\left(1-\frac{1}{n}\right)\frac{3H_{R}^{2}}{8\pi J}.\label{E_SR}
\end{equation}
Because of this, it is impossible to compute the much smaller long-range
energy of Eq. (\ref{E_IM}) by numerical energy minimization.

In the case of random anisotropy, the IM formulas are usually written
in the same way as above with $H_{R}\Rightarrow D_{R}$, which leads
to similar results. The argumentation becomes less transparent because
the uniaxial RA becomes a general tensor after averaging over the
region $R_{f}$. Although the probabilistic estimation for the components
of this tensor remains valid, it is unclear how the spins follow the
averaged RA. It is difficult to argue that they cover the whole sphere
of the possible directions rather than a hemisphere that minimizes
the exchange energy. The result for $R_{f}$ obtained by the GF method
for the 3D Heisenberg RA model reads \citep{prochugar15jmmm_RA}
\begin{equation}
\frac{R_{f}}{a}=15\pi\left(\frac{J}{D_{R}}\right)^{2},\label{Rf_RA}
\end{equation}
where the factor 15 is close to the factor 12 for the 3D Heisenberg
RF model above, Eq. (\ref{Rf_RF}) with $n=3$. For the 2D Heisenberg
model, there is only a numerical result for the magnetic correlation
radius: $R_{f}\simeq11J/D_{R}$ \citep{chugar24prb_sc}.

Using the Imry-Ma argument, one can establish the scaling of different
physical quantities such as, for instance, the microwave absorption
power \citep{chugar24prb_sc}. Instead of performing numerical work
on a system with a weak RA, which has to be big because of the large
$R_{f}$, one can compute a smaller system with a stronger RA and
then obtain the results for the original system with the help of the
scaling. There is a more direct scaling method bypassing the IM argument.
In this method, spin blocks of the size $b$, such as $a\ll b\ll R_{f}$
are introduced \citep{garchu24epl_sc}. Rewriting the original model
on a lattice in the continuous approximation and then discretizing
it again with the new lattice spacing $b$, one obtains the same problem
in terms of a smaller number of spin blocks with a larger ratio $D'_{R}/J$,
which greatly speeds up the computations. Making this scaling transformation,
one obtains the dependence for $R_{f}$ on the parameters which coincides
with the result following from the IM argument. Here, one uses only
the fact that spins are nearly aligned in the region of the linear
size smaller than $R_{f}$, does not assume that the system has zero
magnetization, and does not minimize its energy.

\section{Numerical procedures}

\label{Sec_Numerical-procedures}

The numerical procedures employed here are (i) energy minimization
at $T=0$ and (ii) Metropolis Monte Carlo at $T>0$. The energy minimization
combines aligning the spin $\mathbf{s}_{i}$ with its effective field
$\mathbf{H}_{\mathrm{eff},i}=-\partial\mathcal{H}/\partial\mathbf{s}_{i}$
with the probability $\alpha$ and flipping the spin around the effective
field, $\mathbf{s}_{i}\Rightarrow2\left(\mathbf{s}_{i}\cdot\mathbf{H}_{\mathrm{eff},i}\right)\mathbf{H}_{\mathrm{eff},i}/H_{\mathrm{eff},i}^{2}-\mathbf{s}_{i}$
with the probability $1-\alpha$ (the so-called overrelaxation). The
algorithm uses vectorized updates of columns of spins in checkered
sublattices that allows parallelization of the computation. Thus,
the whole (checkered) spin columns are either aligned with the effective
field or flipped over it. This differs from the non-parallelized version
in which different spins are either aligned or flipped independently,
but the efficiency of both methods is comparable. For the RF model,
overrelaxation is a kind of conservative pseudodynamics, which allows
for better exploration of the phase space of the system. Choosing
a small value of $\alpha$ that has a meaning of the damping in this
procedure (the main choice being $\alpha=0.03$) allows to achieve
a much faster and deeper energy minimization then just the field alignment,
$\alpha=1$. For $\alpha$ smaller than 0.03, the computation becomes
longer again, but the quality of the energy minimization slightly
increases (deeper local energy minima can be found). Overrelaxation
is also very useful for pure systems that have a single energy minimum.

For the model with RA, overrelaxation leads to the energy decrease.
It can be shown that flipping a spin changes the energy by
\begin{equation}
\Delta E=-\frac{D_{R}}{2}\left(\Delta\mathbf{s}_{i}\cdot\mathbf{n}_{i}\right)^{2}.\label{Delta_E_overrelaxation}
\end{equation}
Thus, for RA, the energy minimization can be done with $\alpha=0$,
and there is no control over the rate of the energy decrease, such
as in the RF case.

The Metropolis Monte Carlo routine for classical spin vectors includes
adding a random vector to a spin and normalizing the result to obtain
the trial configuration. After that, the energy change $\Delta E$
is computed and the new spin value is accepted if $\exp\left(-\Delta E/T\right)>\mathrm{rand}$,
where rand is a random number in the interval $\left(0,1\right)$.
If $\Delta E<0$, the trial is automatically accepted. Also here,
updating spins is done in the vectorized and parallelized form for
checkered sublattices. Combining Monte Carlo updates with overrelaxation
greatly speeds up the thermalization. However, for RA systems (as
well as for systems with coherent anisotropy) overrelaxation does
not conserve the energy and it would destroy the detailed balance,
which is the basis of the Monte Carlo method, leading to the energy
lowering. In this case, one should use the method of \textit{thermalized
overrelaxation} \citep{garchu22jpcm} in which a spin is flipped around
the part of $\mathbf{H}_{\mathrm{eff},i}$ that is independent of
the given spin (i.e., the exchange and the applied field). After that,
the energy change $\Delta E$ due to the anisotropy, which can be
positive or negative, is computed, and the usual acceptance/rejection
criterion is applied. Note that flipping the spin over the total effective
field would result in the energy lowering in all cases, and all trials
will be accepted, invalidating the procedure. Thermalized overrelaxation
leads to the correct thermal state, which can be checked by computing
the spin temperature of the system taking into account the single-site
anisotropy \citep{gar21pre}
\begin{equation}
T_{S}=\frac{\sum_{i}\left(\mathbf{s}_{i}\times\mathbf{H}_{\mathrm{eff},i}\right)^{2}}{2\sum_{i,j}J_{ij}\mathbf{s}_{i}\cdot{\bf s}_{j}+D_{R}\sum_{i}\left[3\left(\mathbf{n}_{i}\cdot\mathbf{s}_{i}\right)^{2}-1\right]},\label{TS}
\end{equation}
which for sufficiently large systems at equilibrium is close to the
set temperature $T$.

For computations in a broad region of temperatures across the phase
transition, it is mandatory to employ an adaptive Monte Carlo routine
with an automated stopping criterion, as the number of Monte Carlo
steps (MCS) necessary for equilibrating the system and measuring physical
quantities changes by orders of magnitude. To monitor the equilibration,
the routine uses two control quantities: the energy of the system
and the magnitude $m$ of its magnetization

\begin{figure}
\begin{centering}
\includegraphics[width=8cm]{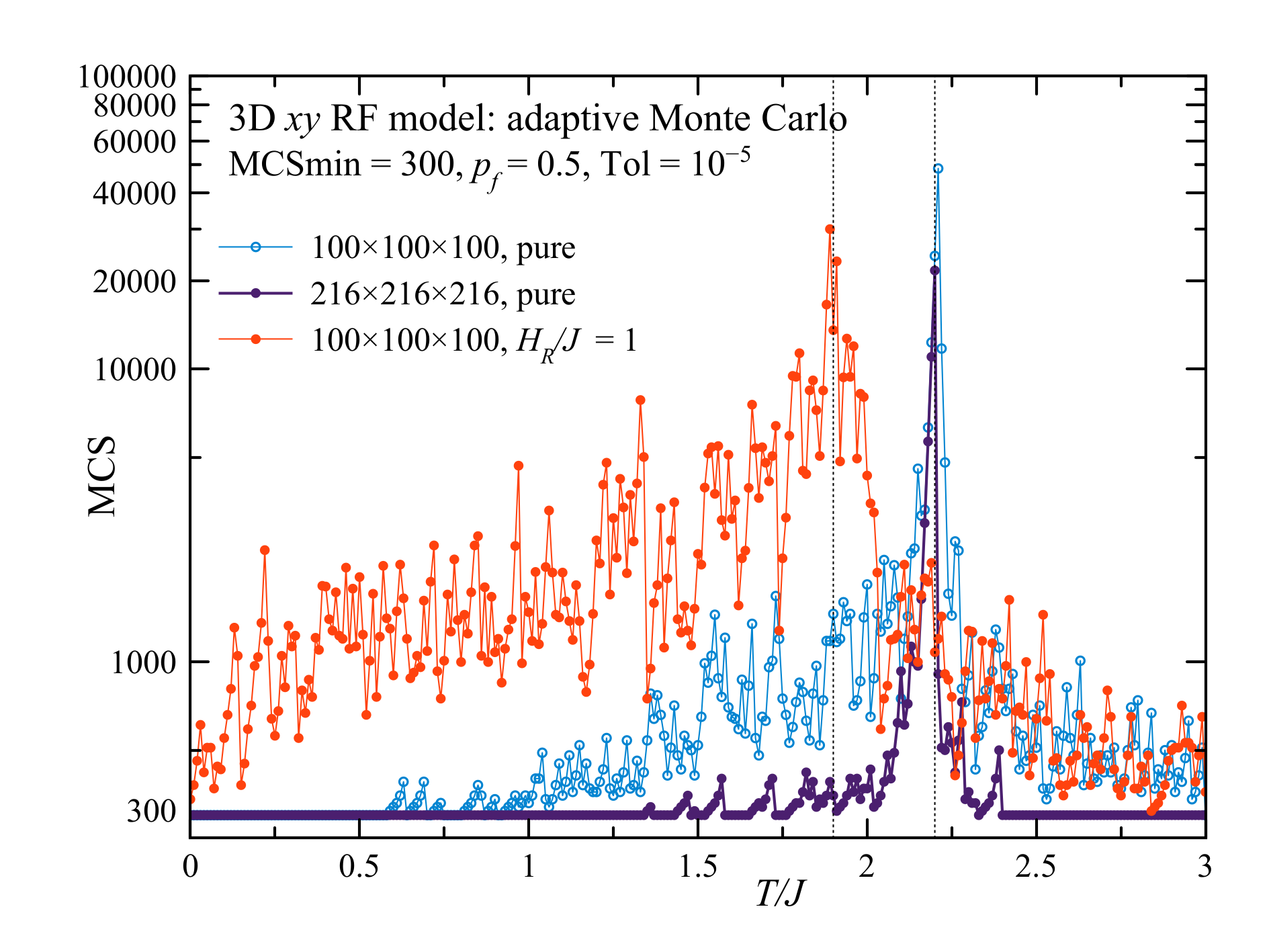}
\par\end{centering}
\caption{Adaptive Monte Carlo: MCS vs $T$ for the 3D $xy$ model.}\label{Fig_MCS_vs_T_RF_xy}
\end{figure}
\begin{equation}
\mathbf{m}=\frac{1}{N}\sum_{i=1}^{N}\mathbf{s}_{i}.
\end{equation}
 The MC simulation is performed in blocks of MCBlock MCS', here $\mathrm{MCBlock}=10$.
After the completion of each block, the control quantities are computed.
The next computational parameter is $\mathrm{MCMass}<1$ (here $\mathrm{MCMass}=0.4$),
which defines the fraction of the computed data used for data processing
and finally for the measurement. The routine forms the lists of the
most recent data of the length $\mathcal{L}=\mathrm{Round}[\mathrm{MCMass}\times N_{\mathrm{val}}]$,
where $N_{\mathrm{val}}$ is the number of values of the physical
quantities computed after each block. Both $N_{\mathrm{val}}$ and
$\mathcal{L}$ increase in the course of the simulation. Then, for
each control quantity, the routine computes the moving average over
its list of length $\mathcal{L}$ using the triangular averaging kernel
$\mathcal{K}_{\mathcal{L},j}$:
\begin{equation}
\mathrm{Q_{mean}}=\sum_{j=1}^{\mathcal{L}}\mathcal{K}_{\mathcal{L},j}\mathrm{QList}_{j},\qquad\sum_{j=1}^{\mathcal{L}}\mathcal{K}_{\mathcal{L},j}=1.\label{Qmean_def}
\end{equation}
The values obtained are put into the list QmeanList of the same length
$\mathcal{L}$. The triangular averaging kernel yields a rather smooth
list of data, which is used to quantify the evolution of the control
quantities. As $K_{\mathcal{L},j}$ has a maximum at $j=\mathcal{L}/2$,
the values from the middle of the averaging interval $\mathcal{L}$
make the largest contribution.

\begin{figure}
\begin{centering}
\includegraphics[width=8cm]{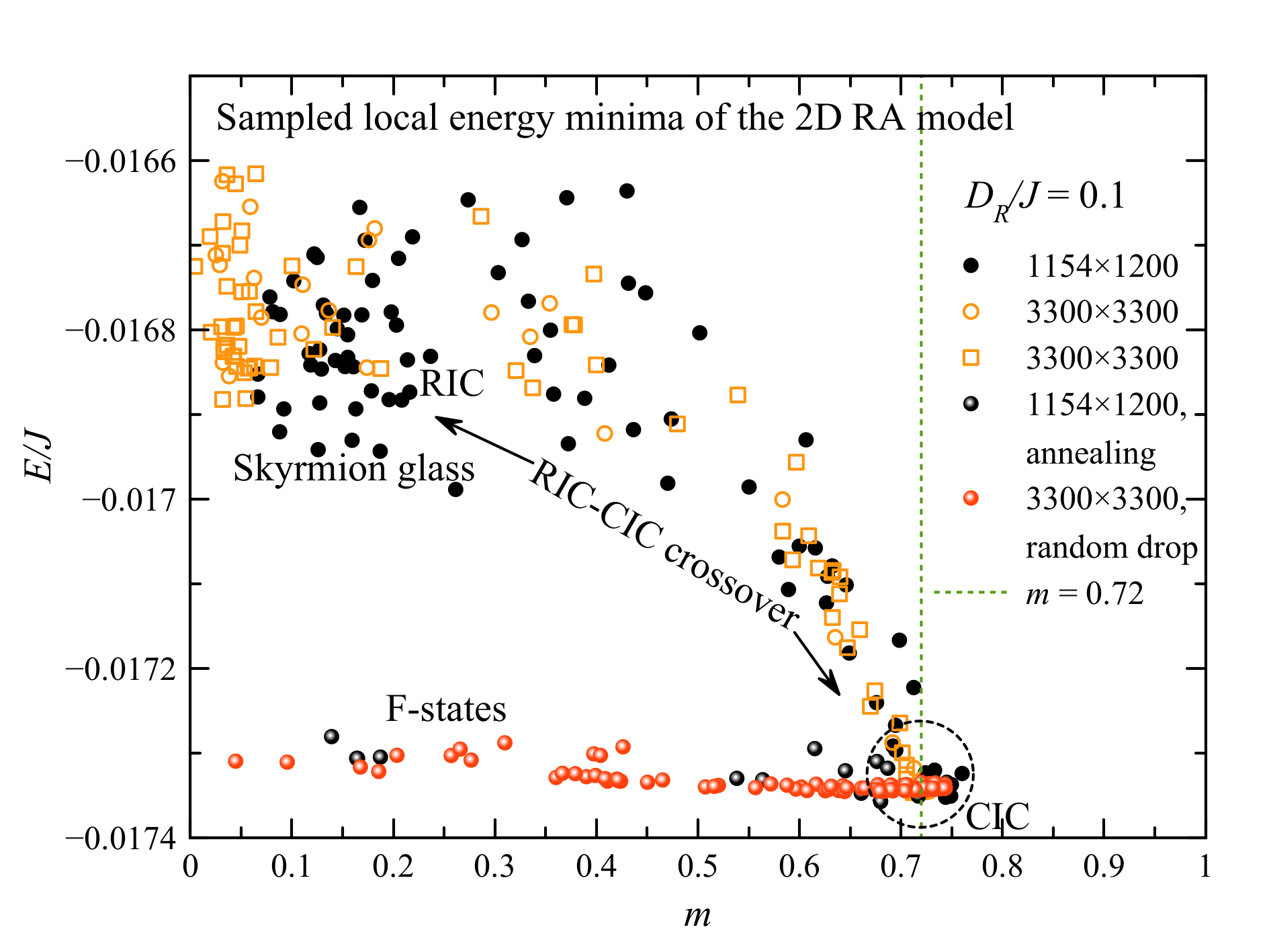}
\par\end{centering}
\begin{centering}
\includegraphics[width=8cm]{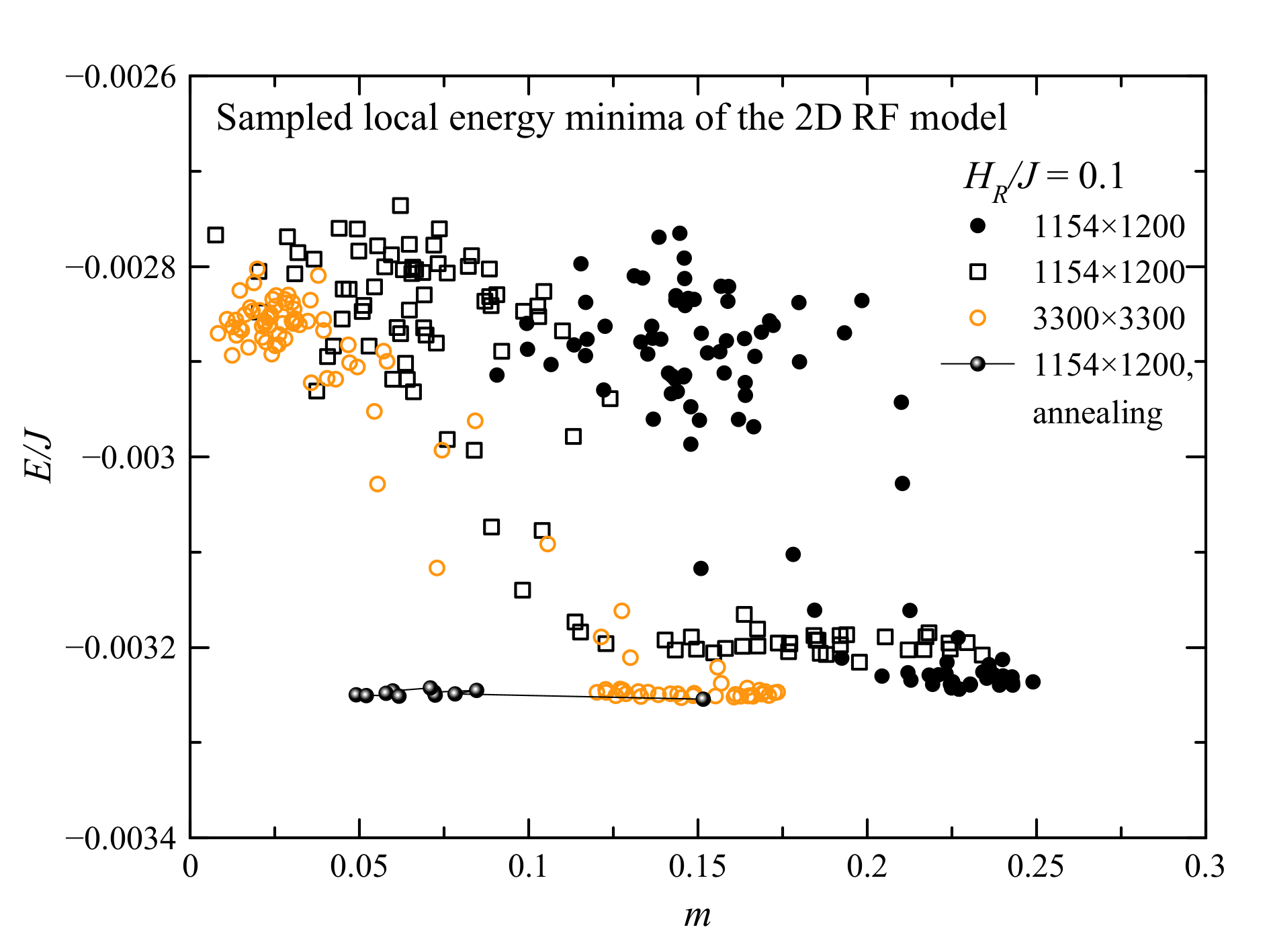}
\par\end{centering}
\caption{Local energy minima of 2D models with quenched randomness sampled
by the RIC-CIC crossover, random dropping, and simulated annealing.
Upper panel: 2D RA model. Lower panel: 2D RF model. }\label{Fig_E_vs_m_2D}
\end{figure}

\begin{figure}
\begin{centering}
\includegraphics[width=8cm]{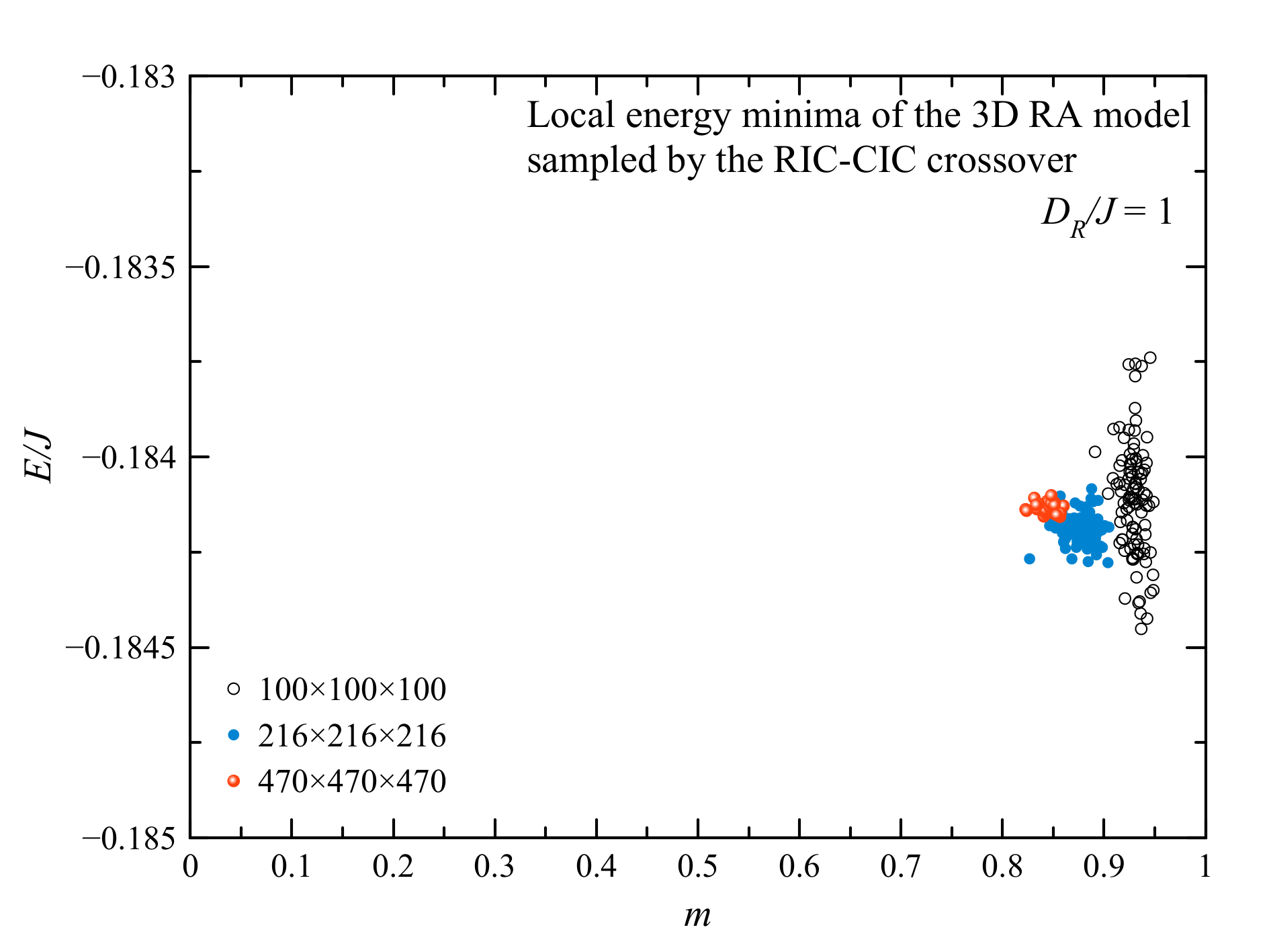}
\par\end{centering}
\begin{centering}
\includegraphics[width=8cm]{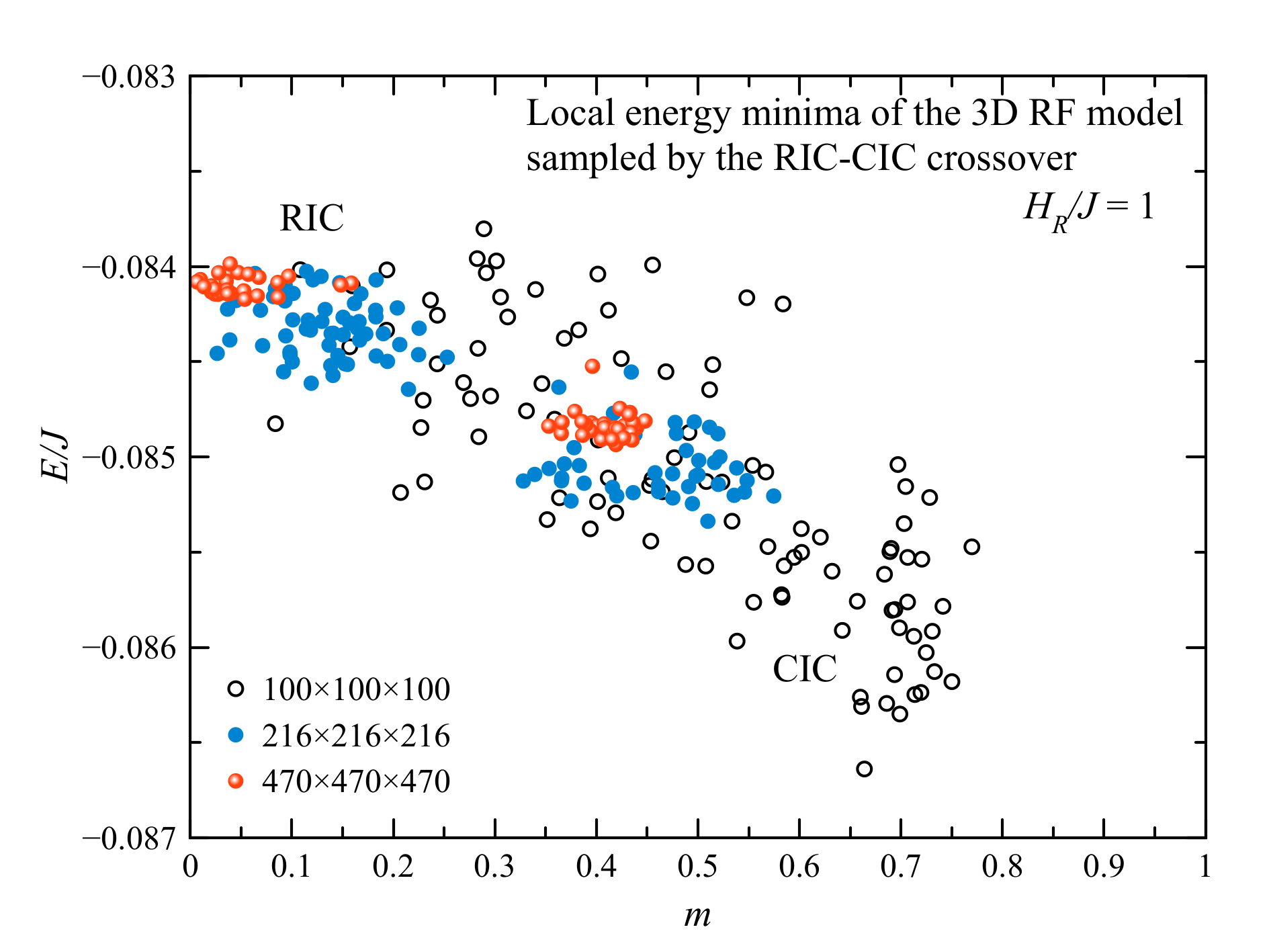}
\par\end{centering}
\caption{Local energy minima of 3D models with quenched randomness sampled
by the RIC-CIC crossover. Upper panel: 3D RA model. Lower panel: 3D
RF model. States on the left with lower magnetization and higher energies
contain singularities (hedgehogs). }\label{Fig_E_vs_m_CIC_RIC_3d_3comp}
\end{figure}
\begin{figure}
\begin{centering}
\includegraphics[width=8cm]{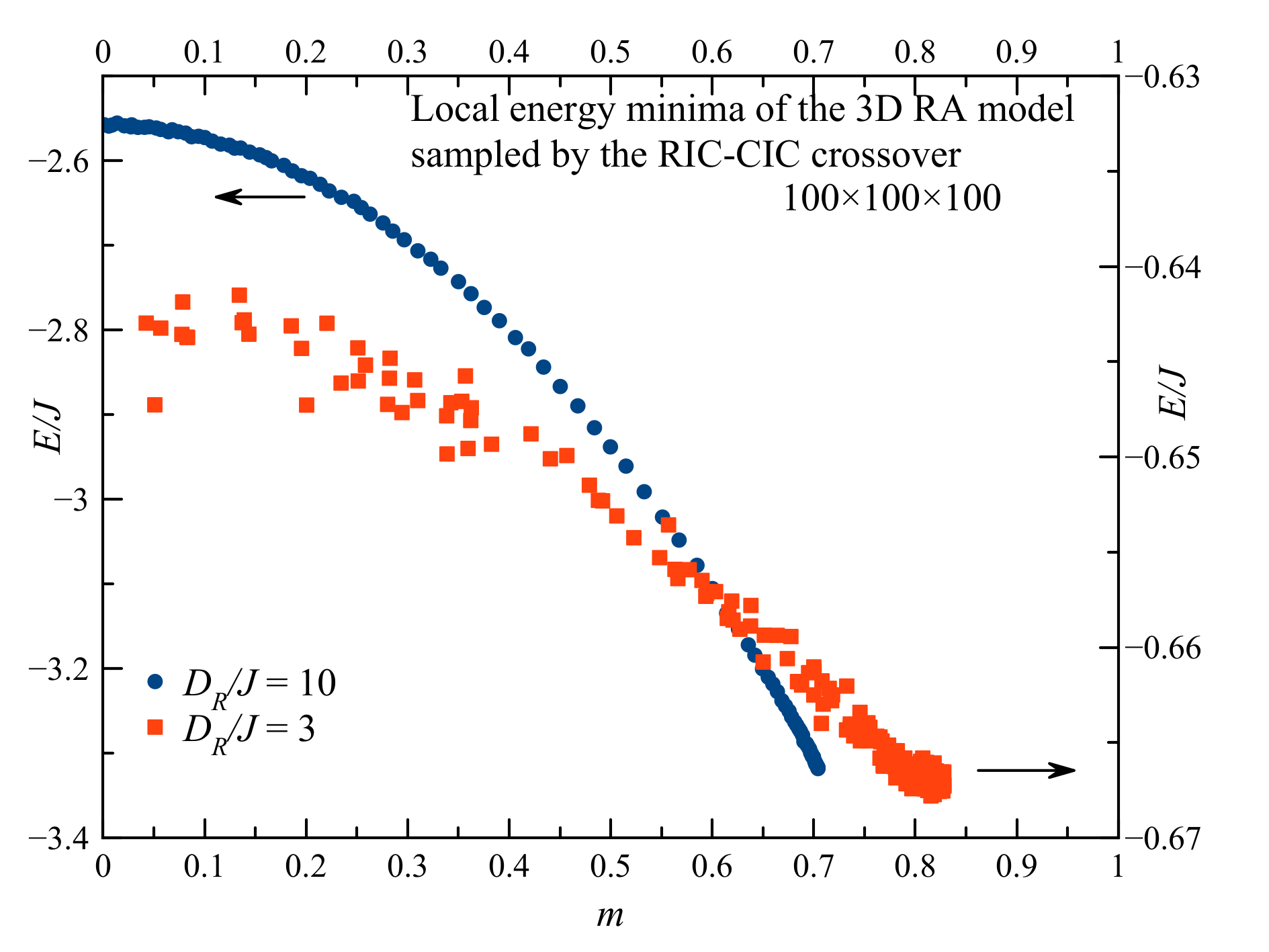}
\par\end{centering}
\caption{Local energy minima of 3D models with quenched randomness sampled
by the RIC-CIC crossover: $D_{R}/J=3$ and 10. Energy minima with
smaller magnetization and higher energy contain singularities (hedgehogs).
}\label{Fig_E_vs_m_CIC_RIC_3D_3comp_DR=00003D3_10}

\end{figure}

Next, the average slope of the QmeanList per one block is defined
as
\begin{equation}
\mathrm{Q}_{\mathrm{slope}}=\frac{4}{\mathcal{L}^{2}}\sum_{j=1}^{\mathcal{L}}\left|\mathrm{QmeanList}_{j}-\left\langle \mathrm{QmeanList}\right\rangle \right|.\label{CQSlope_def}
\end{equation}
A sufficiently small slope could be used as the stopping criterion
for the thermalization. However, for additional focusing on the regions
where thermalization is slow, the routine uses the modified slope
\begin{equation}
\mathrm{\tilde{Q}}_{\mathrm{slope}}=\mathrm{Q}_{\mathrm{slope}}\left(\frac{\mathrm{MCS}}{\mathrm{MCSmin}}\right)^{p_{f}},\label{SlopePar_def}
\end{equation}
where MCS is the total number of MCS from the beginning of the simulation
and MCSmin is the minimal number of MCS required (here $\mathrm{MCSmin}=300$).
The power $p_{f}$ controls the focusing: no focusing for $p_{f}=0$
and strong focusing for $p_{f}=0.5$ (here 0.35 and 0.5). One can
say that focusing effectively decreases the tolerance Tol if the equilibration
is slow. The stopping criterion for the equilibration reads
\begin{equation}
\mathrm{\tilde{Q}}_{\mathrm{slope}}<Q_{0}\mathrm{\times MCBlock}\times\mathrm{Tol,}\label{StoppingCriterion}
\end{equation}
where $Q_{0}$ is the characteristic value of the quantity $Q$ ($J$
for the energy and 1 for the magnetization) and Tol is the tolerance,
here mostly $\mathrm{Tol}=10^{-5}$. For stopping, this criterion
should be fulfilled for all control quantities, and MCS should exceed
MCSmin. The final values of the control quantities and all other physical
quantities are obtained via averaging of the most recent portion of
the data of length $\mathcal{L}$ with the averaging kernel $\mathcal{K}$,
as in Eq. (\ref{Qmean_def}). In this approach, thermalization and
measurement are not separated.

Computations were performed with Wolfram Mathematica in the vectorized,
parallelized, and compiled form on different workstations with 16-24
processor cores and 32-64 GB of memory. The maximal system size is
$10^{8}$ spins for 2D and 3D systems.

Figure \ref{Fig_MCS_vs_T_RF_xy} illustrates the adaptive Monte Carlo
method for the 3D $xy$ model, computed with $\mathrm{MCSmin}=300$,
$p_{f}=0.5$, and $\mathrm{Tol}=10^{-5}$. The number of MSC done
before stopping reaches its maximum at the phase transition point:
the Curie temperature $T/J=2.2$ in the pure model and the correlated
spin-glass transition temperature $T/J=1.9$ for $H_{R}/J=1$. For
the larger system of $216\times216\times216$ spins, $\mathrm{MCS=MCSmin}$
far from the transition point, whereas for the smaller pure system
of $100\times100\times100$ spins $\mathrm{MCS>MCSmin}$ in this region
because of stronger fluctuations in smaller systems. In the glassy
system with $H_{R}/J=1$, equilibration is more difficult and requires
more MCS. However, in these systems, the equilibrium is not well defined
at low temperatures and cannot be reached in numerical experiments.

\section{Numerical results}

\label{Sec_Numerical-results}

\subsection{RIC-CIC crossover and other methods of sampling local energy minima}

\label{Sec_RIC-CIC}

Large random-field and random-anisotropy systems have a huge number
of local energy minima. Although it is difficult to find the ground
state and enumerate all the local energy minima, one can get an idea
of the energy minima by sampling them. There are several methods of
sampling, one of them being the RIC-CIC crossover. Here, the energy
minimization at $T=0$ is performed starting from the combined initial
condition
\begin{equation}
\mathbf{s}_{\mathrm{inilial}}=\mathrm{Normalize}\left[\left(1-\xi\right)\mathbf{s}_{\mathrm{RIC}}+\xi\mathbf{s}_{\mathrm{CIC}}\right],
\end{equation}
where $\mathbf{s}_{\mathrm{initial}}$ is a tensor containing all
spins in the system, $\mathbf{s}_{\mathrm{RIC}}$ is that with randomly
directed spins, $\mathbf{s}_{\mathrm{CIC}}$ is that with collinear
spins, $0<\xi<1$ is a mixing parameter, and the result of mixing
the RIC and CIC states is normalized by one on each lattice site.
For $\xi$ close to zero, there is a greater possibility that the
resulting state is rich in singularities or other topological structures
and has a small magnetization. For $\xi$ close to one, the resulting
state should have fewer or no singularities and a larger magnetization.
Energy minimization was performed for many values of $\xi$ with different
realizations of the RA.

Other methods of sampling local energy minima are so-called \textit{simulated
annealing} and \textit{random dropping}. Both methods sample energy
minima with significant magnetization and lower energy, the so-called
F-states \citep{garchupro13prb,garchu15epjb}. Simulated annealing
is a numerical experiment in which the magnetic field $H$ is swept
back and forth at $T=0$ with a decreasing amplitude. The values of
the energy and magnetization are recorded at all points where $H=0$.
The random-dropping experiment consists of first minimizing the energy
at $T=0$ with the magnetic field $H$ applied in a random direction
and then switching the field off and minimizing the energy again (dropping).
The value of $H$ is chosen to be significant but still far from saturation.

Local energy minima of a 2D 3-component RA model sampled by the RIC-CIC
crossover, random dropping, and simulated annealing are shown in the
upper panel of Fig. \ref{Fig_E_vs_m_2D}. The skyrmion glass \citep{chugar18prl,chugar18njp}
obtained for $\xi\ll1$ has a higher energy and smaller magnetization
than the F-states, which have their lowest energy near $m\simeq0.72$,
obtained with $\xi\simeq1$. Random-dropping and simulated annealing
experiments yield a broader manifold of F-states with lower magnetization
and slightly higher energy (red and black balls in the figure). This
can be explained by the bending of the magnetized state, which, over
large distances, can significantly reduce the magnetization while
only slightly increasing the energy. These results indicate that the
2D RA model has an ordered ground state, contrary to the predictions
based on the IM argument. However, this system does not order on lowering
the temperature. One of the reasons is that skyrmions present at elevated
temperatures do not disappear with lowering $T$ because they are
topologically protected and pinned by the RA. Also, the exchange connectivity
of a 2D system with a continuous symmetry is insufficient for ordering.
Results similar to those in Fig. \ref{Fig_E_vs_m_2D} are shown in
Fig. 16 of Ref. \citep{garchupro13prb} for the 3D $xy$ RF system.

\begin{figure}
\begin{centering}
\includegraphics[width=8cm]{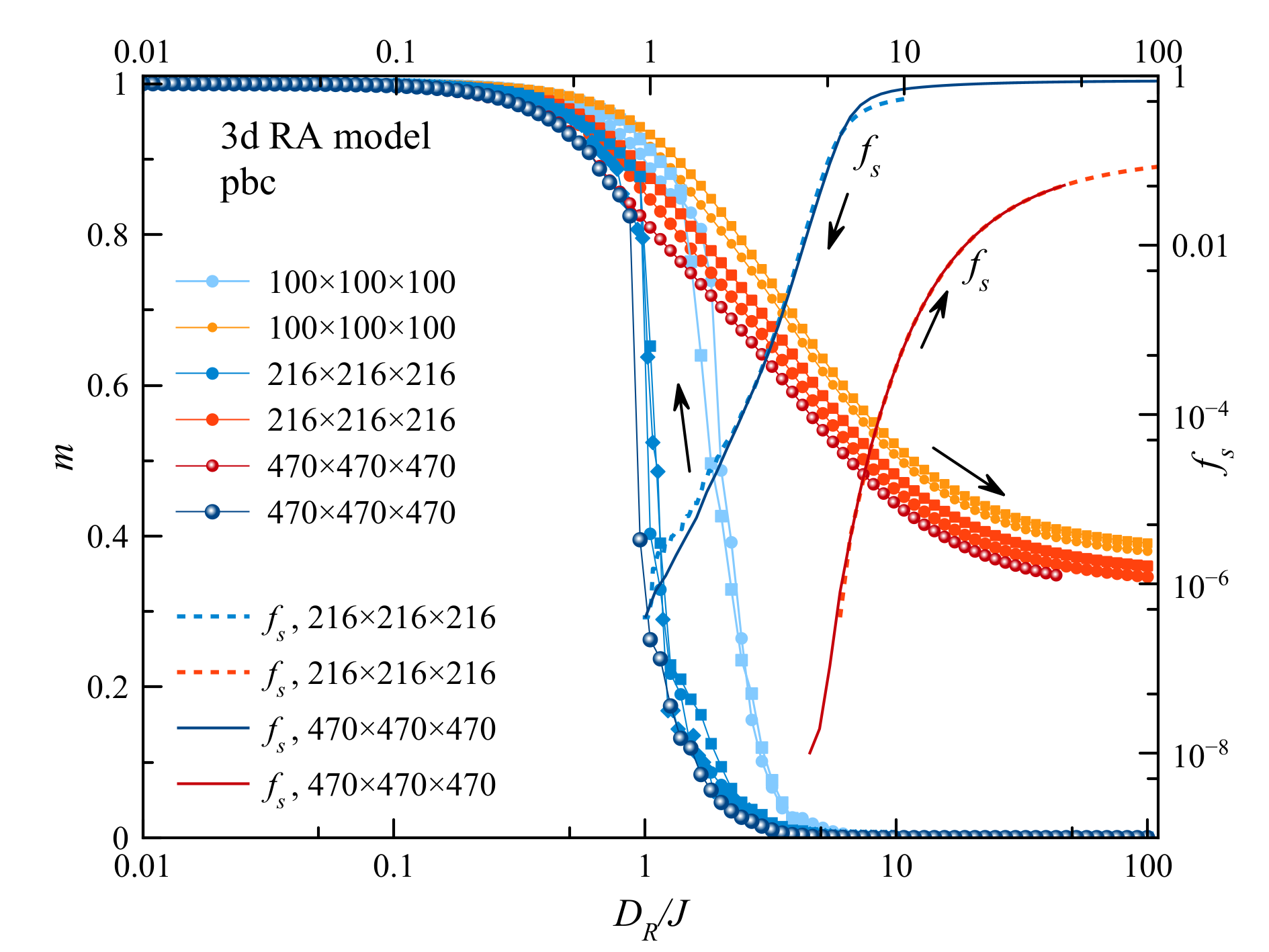}
\par\end{centering}
\begin{centering}
\includegraphics[width=8cm]{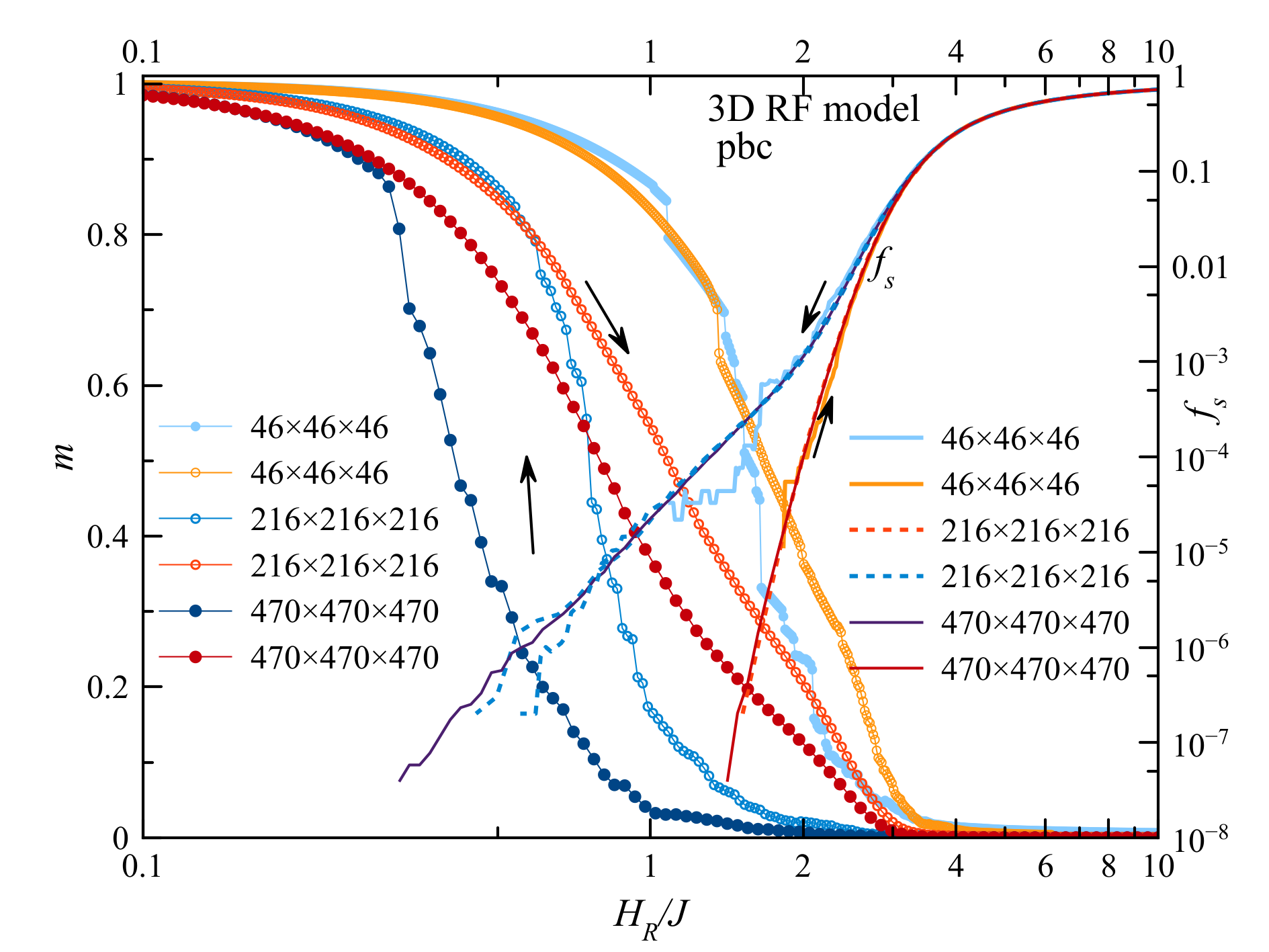}
\par\end{centering}
\caption{$D_{R}$ and $H_{R}$ scans with energy minimization of 3D Heisenberg
systems at $T=0$. Upper panel: RA model. Lower panel: RF model.}\label{Fig_DR_and_HR_scans}
\end{figure}

\begin{figure}
\begin{centering}
\includegraphics[width=8cm]{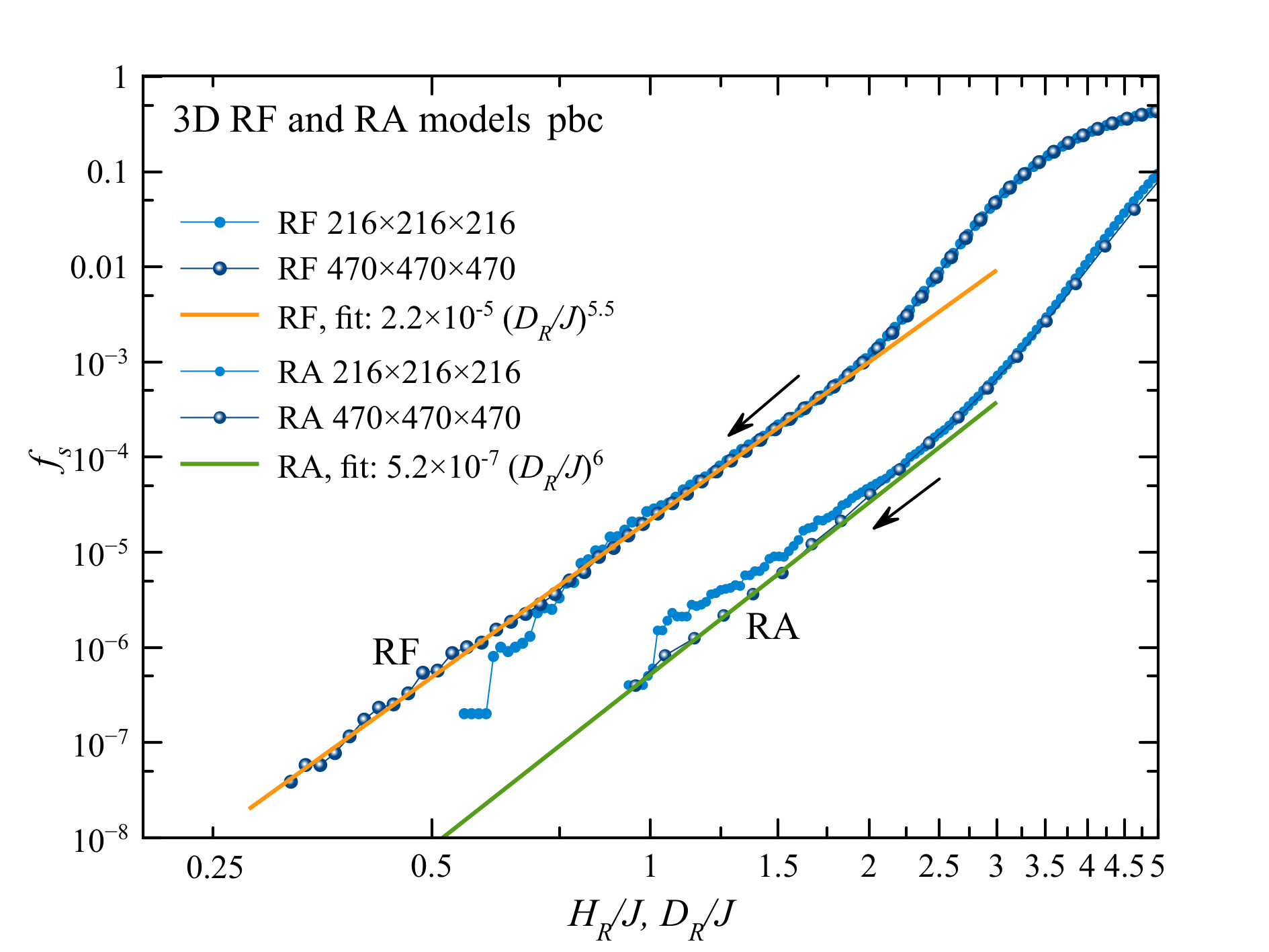}
\par\end{centering}
\caption{Concentration of hedgehogs in the 3D Heisenberg RA and RF models on
lowering $D_{R}$ and $H_{R}$. }\label{Fig_fs_vs_HR_DR}
\end{figure}

The lower panel of Fig. \ref{Fig_E_vs_m_2D} shows the sampled local
energy minima of the 2D RF model. The difference is that in the RA
model, the ground state has a large magnetization and does not depend
on the system size, whereas in the RF model, the magnetization in
the ground state is small and decreases with the system size. One
can expect its vanishing in the large-system limit.

Local energy minima of 3D Heisenberg RF and RA models for $D_{R}/J=1$
and $H_{R}/J=1$, sampled by the RIC-CIC crossover method, are shown
in Fig. \ref{Fig_E_vs_m_CIC_RIC_3d_3comp}. For the RA model (upper
panel of Fig. \ref{Fig_E_vs_m_CIC_RIC_3d_3comp}), energy minimization
ends up in the states with a large magnetization and no singularities
independently of the initial condition. Increasing the system size
lowers the magnetization in the ground state only insignificantly.
This is a clear demonstration of a strong ordering tendency in the
3D RA model, contrary to the usually made statement that the RA destroys
the ordered state in this case. For larger $D_{R}/J$, the distribution
of the local energy minima is broader because there are states with
singularities having smaller magnetization and larger energy, which
are obtained by the energy minimization starting from the states closer
to RIC. The results for $D_{R}/J=3$ and 10 are shown in Fig. \ref{Fig_E_vs_m_CIC_RIC_3D_3comp_DR=00003D3_10}.

In the case of the RF model (lower panel of Fig. \ref{Fig_E_vs_m_CIC_RIC_3d_3comp}),
the states with higher magnetization have lower energy, although the
magnetization is smaller than for the RA model and decreases with
the system size. These states originate from the energy minimization
starting from the initial states close to CIC and have no singularities.
The states with smaller magnetization originate from the initial states
close to RIC, and they have singularities (hedgehogs or Bloch points
\citep{doe68jap,kuchkinetal25prr} for the 3D Heisenberg model) that
increase their energy. The role of singularities will be investigated
in the next section.

\begin{figure}
\begin{centering}
\includegraphics[width=8cm]{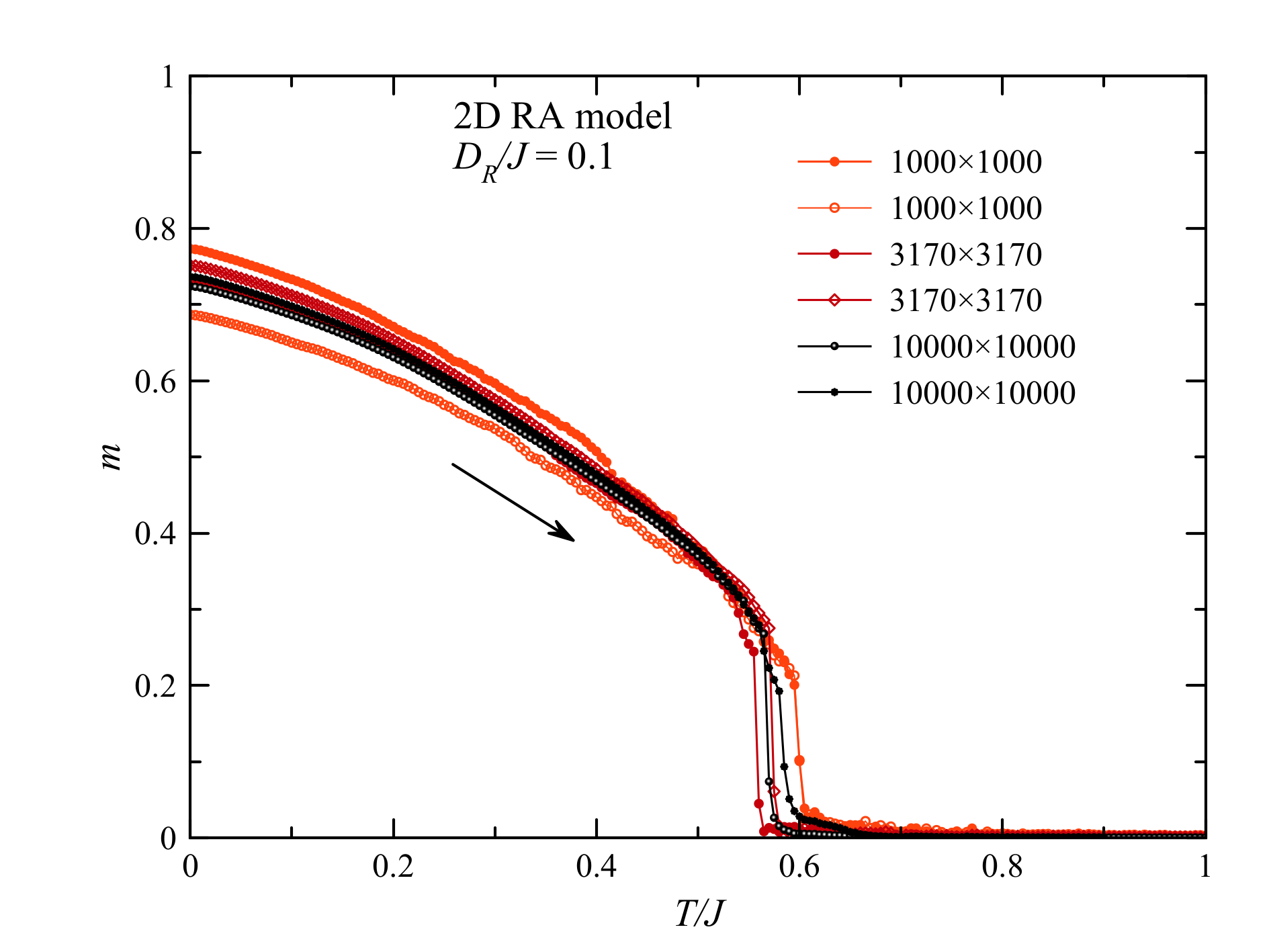}
\par\end{centering}
\caption{The (increasing) temperature dependence of the magnetization in the
2D RA model. }\label{Fig m_vs_T_2D_DR=00003D0.1}
\end{figure}

\begin{figure}
\begin{centering}
\includegraphics[width=8cm]{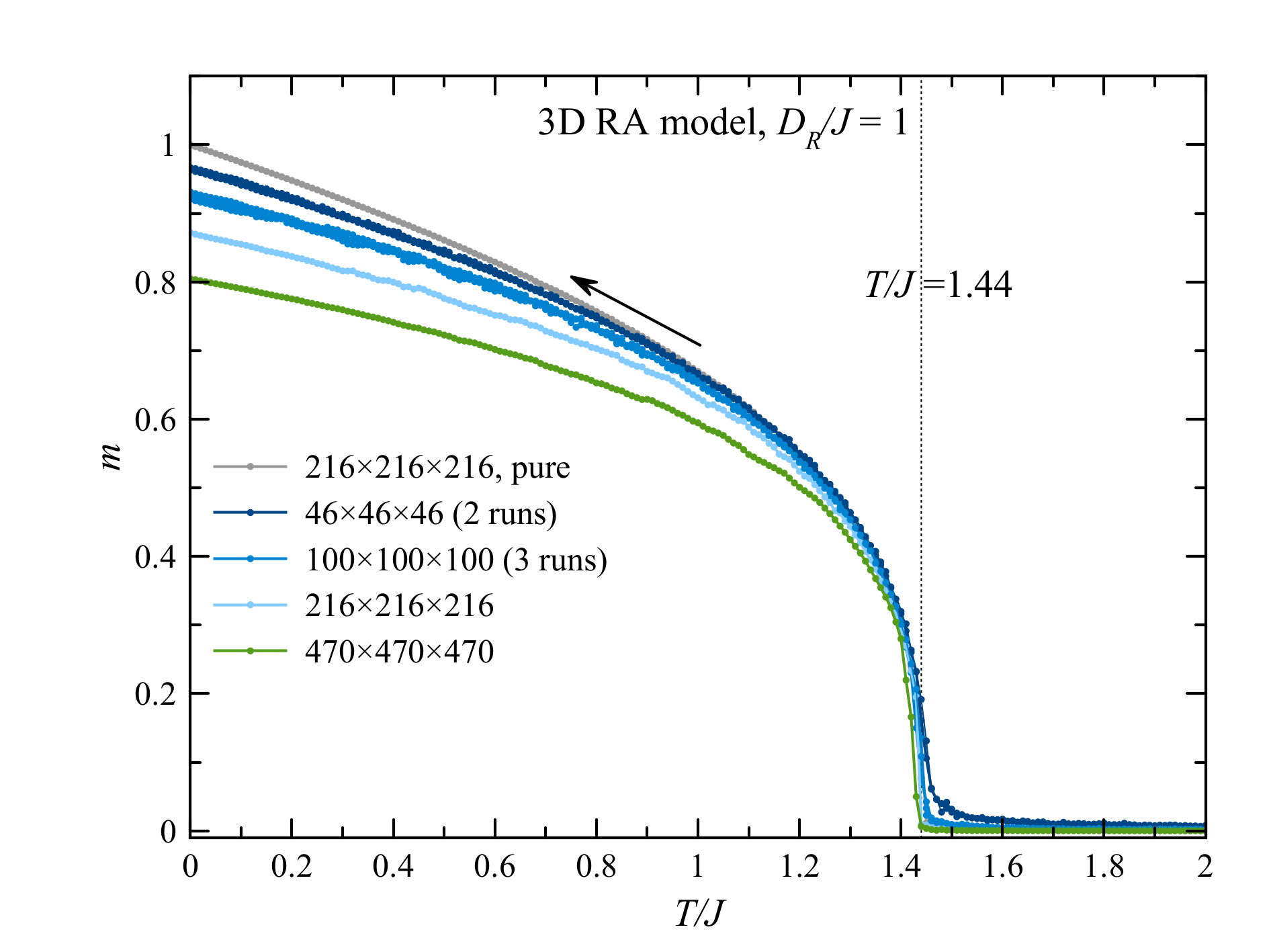}
\par\end{centering}
\begin{centering}
\includegraphics[width=8cm]{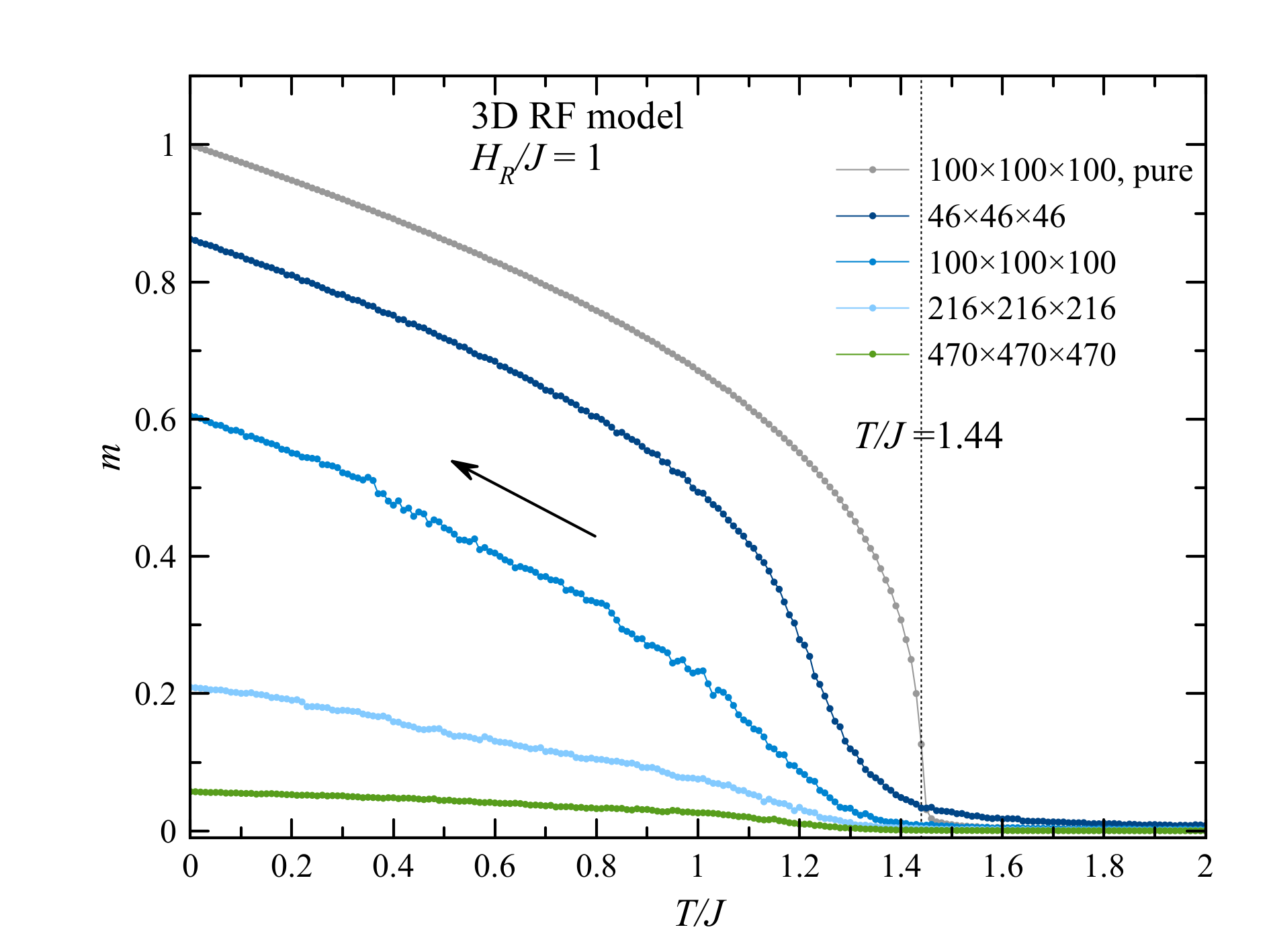}
\par\end{centering}
\caption{Temperature dependences of the magnetization of the 3D Heisenberg
model with RA and RF obtained on lowering $T$. Upper panel: RA model.
Lower panel: RF model. }\label{Fig_m_vs_T_3D}

\end{figure}
\begin{figure}
\begin{centering}
\includegraphics[width=8cm]{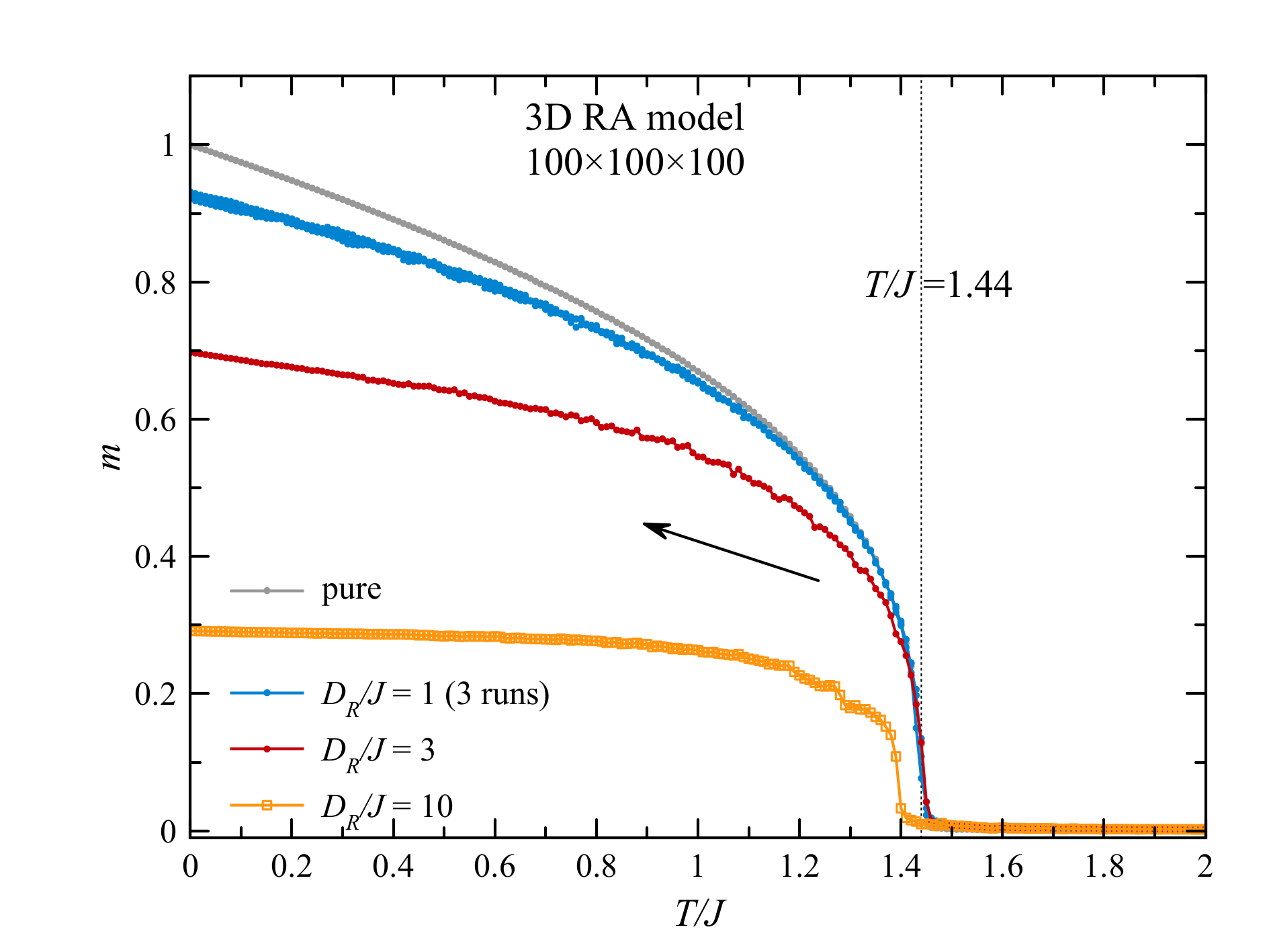}
\par\end{centering}
\caption{Temperature dependences of the magnetization of the 3D RA Heisenberg
model with different values of the RA strength $D_{R}$.  }\label{Fig_m_vs_T_100x100x100_DR_var}

\end{figure}

\subsection{$D_{R}$ and $H_{R}$ scans at $T=0$}

\label{Sec_DR_and_HR_scans}

The second type of numerical experiments performed here is $D_{R}$
and $H_{R}$ scans on the systems with the same RA or RF realization,
with the energy minimization at $T=0$. The values of $D_{R}$ and
$H_{R}$ are increased or decreased by a small amount each time, so
that the initial state of the system is already close to its final
state and the energy minimization is fast. The dependences of the
magnetization $m$ and other quantities are different for increasing
and decreasing the strength of RA or RF, as can be seen in Fig. \ref{Fig_DR_and_HR_scans}.
When the strength of the RA or RF increases (red curves), the magnetization
starts at $m=1$ for small $D_{R}/J$ or $H_{R}/J$ on the left side
of the figures, then it decreases because if the progressive disordering.
For the RA model, this ends up in a partial disordering at large $D_{R}/J$,
which is roughly consistent with the spins directed within a hemisphere,
$m=1/2$. In fact, the disordering is somewhat stronger with $m$
going down to $m\simeq0.3$, which means the system in not exactly
in its ground state (the hemisphere distribution of spins) in the
limit $D_{R}\rightarrow\infty$, which is an effect of this particular
kind of evolution. For the RF model, the magnetization vanishes in
the limit $H_{R}\rightarrow\infty$, as expected.

In this experiment, the concentration of singularities (hedgehogs)
$f_{S}$ was also computed. In the absence of a topological formula,
the sum of 8 spins surrounding the center of each unit cell was computed.
If the average of these 8 spins had a magnitude smaller than 0.5,
a hedgehog was detected. The concentration of hedgehogs was computed
as the number of hedgehogs divided by the number of cells in the system.
On increasing $D_{R}$ or $H_{R}$, there are no singularities at
the beginning, but they emerge at critical values of $D_{R}$ or $H_{R}$,
then $f_{S}$ grows quickly.

\begin{figure}
\begin{centering}
\includegraphics[width=8cm]{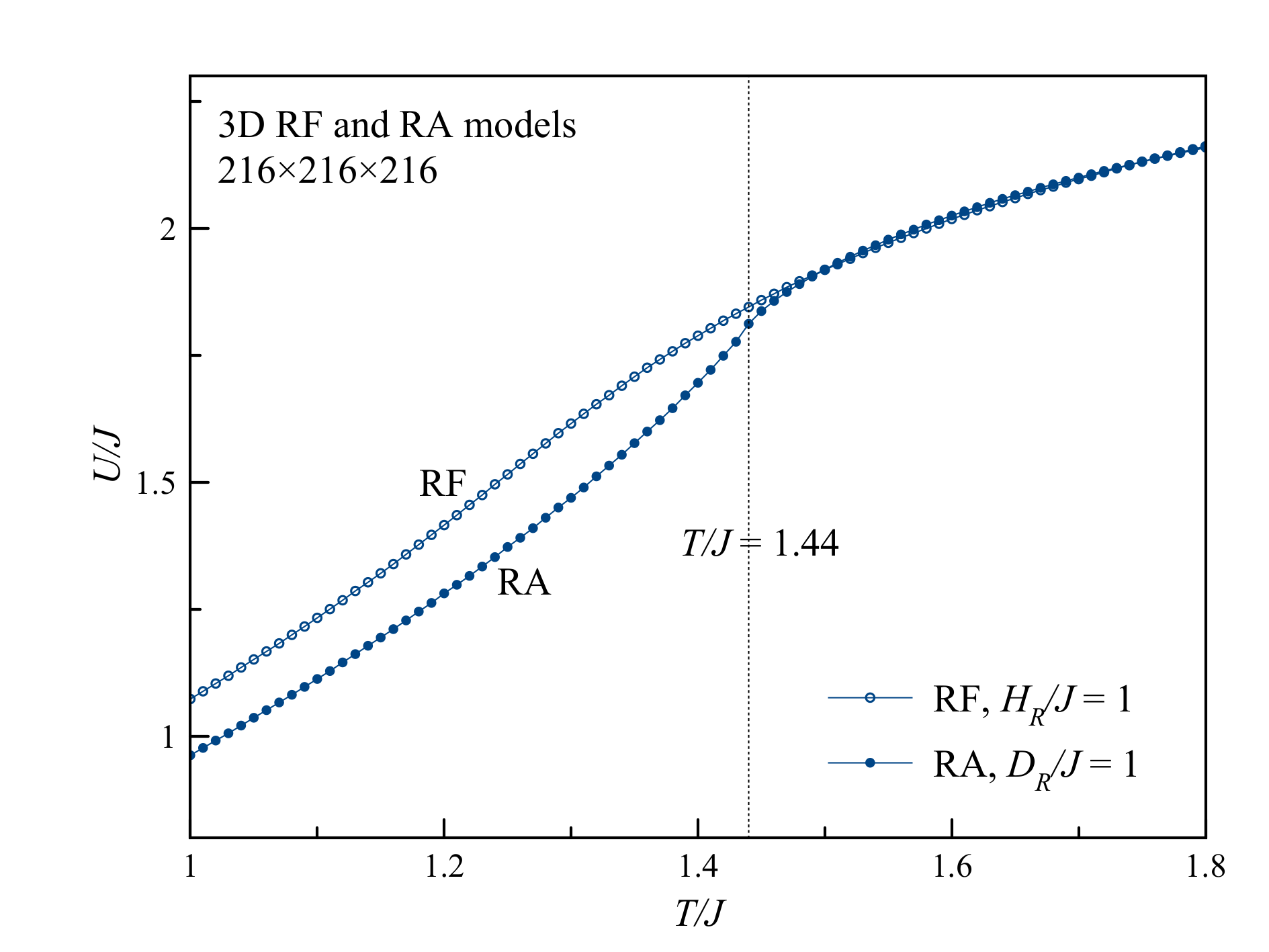}
\par\end{centering}
\caption{Temperature dependence of the energy of the 3D RA and RF models. }\label{Fig_U_vs_T}

\end{figure}
\begin{figure}
\begin{centering}
\includegraphics[width=8cm]{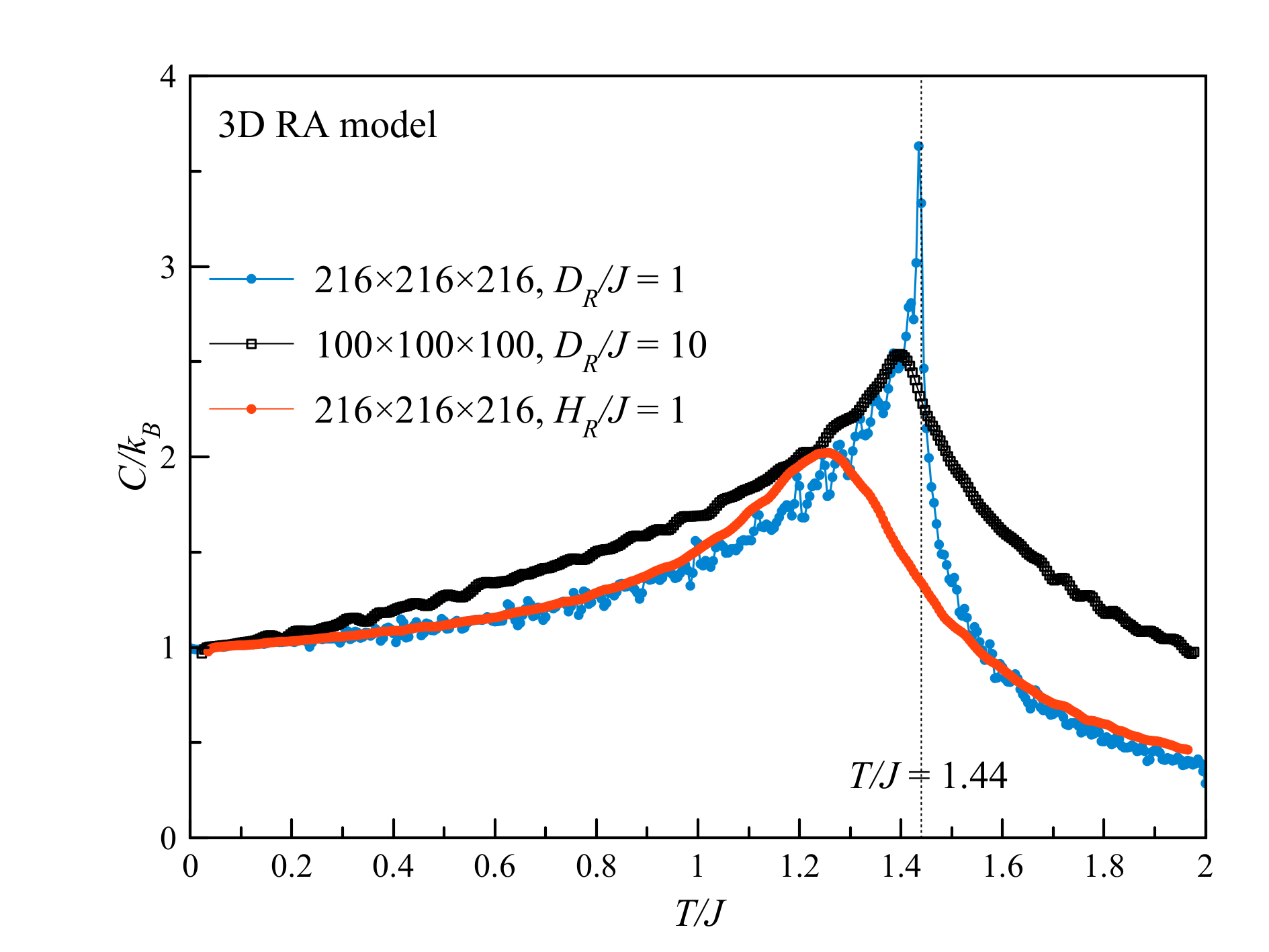}
\par\end{centering}
\caption{Temperature dependences of the heat capacity of the 3D RA and RF models.
}\label{Fig_C_vs_T}

\end{figure}

On decreasing $D_{R}$ or $H_{R}$ from large values to zero, there
are lots of singularities at the beginning, and, accordingly, $m$
is close to zero. With decreasing $D_{R}$ or $H_{R}$, the exchange
suppresses more and more singularities which eventually leads to nonzero
values of $m$. In Fig. \ref{Fig_DR_and_HR_scans}, one can see a
clear correlation between the concentration of singularities and the
magnetization. In both models, when $f_{s}=0$ is reached, $m$ strongly
increases and then continues to increase smoothly as $D_{R}$ or $H_{R}$
are further lowered. The difference between the two models is that
in the case of RA the critical value of $D_{R}$ at which $f_{s}$
vanishes becomes independent of the system size at large sizes such
as $216^{3}\simeq10^{7}$ spins and as $470^{3}\simeq10^{8}$ spins,
whereas in the RF case singularities turn out to be more resilient
and disappear at lower $H_{R}$ for larger system size. That is, for
the RA, there exists a well-defined critical value of $D_{R}/J$ below
which singularities cannot be supported and are suppressed by the
exchange. For the RF, it appears that singularities exist for whatever
small values of $H_{R}$ and their disappearance in systems of finite
size is due to a too small concentration of singularities. Accordingly,
the magnetization curves in the lower panel of Fig. \ref{Fig_DR_and_HR_scans}
shift to the left with the increase of the system size.

The dependence of the concentration of singularities on the lowered
$D_{R}$ and $H_{R}$ is shown in the separate Fig. \ref{Fig_fs_vs_HR_DR}.
Whereas for the RA model, $f_{s}=0$ is reached at $D_{R}/J\simeq1$
for both system sizes, for the RF model there are many more singularities
with no well-defined critical value of $H_{R}$ at which they disappear.

Applying the topological argument of Refs. \citep{garchupro13prb,progarchu14prl}
to the RA model, one can relate singularities and magnetic ordering.
Any fully random configuration of $n$-component spins in $d$ dimensions
contains singularities if $n\leq d$. Particularly for the 3D Heisenberg
model ($n=3$) there are 2D subspaces with $s_{x}=0$, $s_{y}=0$,
and $s_{z}=0$. All three subspaces cross at some points where all
three components of the spin field are zero. This means there are
singularities at these points. Spin configurations containing singularities
have zero magnetization in the limit of a large system size. If for
$n\leq d$ there are no singularities in the system, disordering of
spins is not complete, and there is a nonzero magnetization for large
system sizes. This is the case for the 3D RA model for $D_{R}/J\leq1$
where the singularities are suppressed by the strong exchange.

One can perform a similar investigation of 2D RA and RF models. One
result is Fig. 3 of Ref. \citep{chugar24prb_sc} for the 2D RA model
obtained by the energy minimization from the CIC for all values of
$D_{R}/J$. This result has an interesting and so far unexplained
feature -- a nearly constant value of the magnetization $m\simeq0.715$
in a broad interval of $D_{R}/J$, which is followed by a decrease
starting at $D_{R}/J\simeq6$. A systematic investigation of the $D_{R}$
and $H_{R}$ scans for 2D models is not performed here to save space.

\begin{figure}
\begin{centering}
\includegraphics[width=8cm]{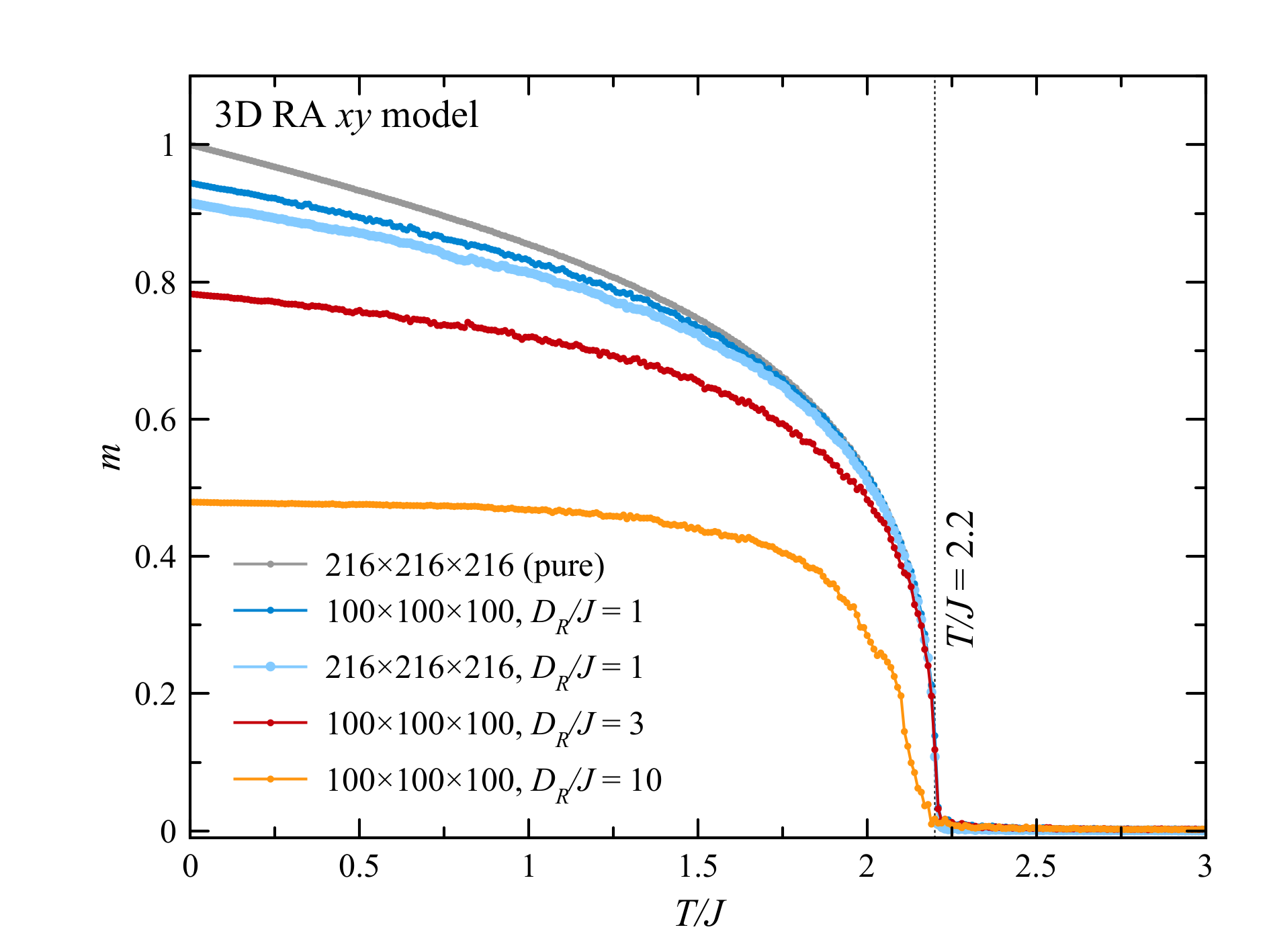}
\par\end{centering}
\begin{centering}
\includegraphics[width=8cm]{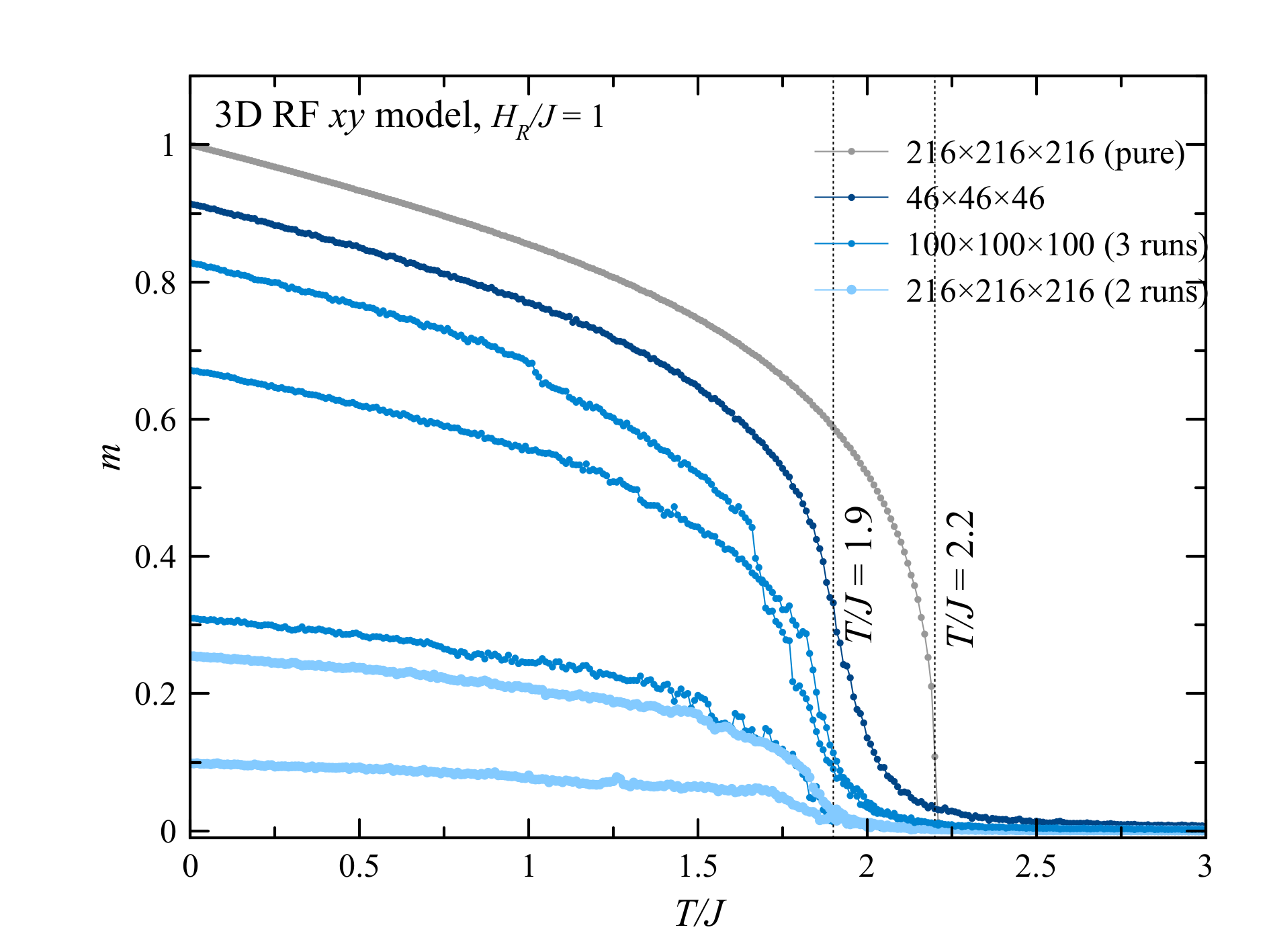}
\par\end{centering}
\caption{Temperature dependence of the magnetization of the 3D $xy$ RA and
RF models. Upper panel: RA model. Lower panel: RF model. }\label{Fig_m_vs_T_3D_xy}

\end{figure}

\subsection{Temperature dependences}

Finally, the temperature dependences of the physical quantities were
computed. The 2D RA model was investigated in Ref. \citep{garchu22jpcm}.
On lowering the temperature, this system of a sufficiently large size
does not order. The question was whether it freezes into a correlated
spin glass. It was found that on lowering the temperature, spins stop
rotating because of blocking by the energy barriers created by the
RA rather than by spin-glass freezing as a many-body effect. The heat
capacity of this system has a rounded maximum without a cusp characteristic
for spin glasses. Here, the question of the thermal stability of the
2D RA system prepared in the ordered state is investigated. It is
well known that in the pure 2D Heisenberg ferromagnet, magnetic order
is destroyed by thermal fluctuations. Monte Carlo simulations give
the magnetization $m(T)$, which steadily decreases with the system
size, and there is no well-defined transition temperature. How stable
is the magnetized state of the 2D RA model at nonzero temperatures?
The results of the Monte Carlo simulations using the Metropolis sampling
combined with the thermalized overrelaxation are shown in Fig. \ref{Fig m_vs_T_2D_DR=00003D0.1}.
At the beginning, the energy was minimized at $T=0$ starting with
CIC, then the temperature was increased in small steps, and the Monte
Carlo equilibration was performed. Fig. \ref{Fig m_vs_T_2D_DR=00003D0.1}
shows that the magnetized state is thermally stable. There are differences
in the $m(T)$ curves due to different realizations of the RA, but
they decrease with the system's size. The magnetization decreases
with the temperature and then disappears in a jump. This behavior
can be explained by the blocking mechanism. There are correlated regions,
the Imry-Ma domains, pinned in the magnetized state by the anisotropy
barriers. With increasing $T$, the magnetization in the IM domains
decreases, which is seen in the figure. Finally, collective spins
of the correlated regions unblock and start flipping over the barriers,
which leads to the demagnetized state, $m=0$. The sharp unblocking
transition is the result of the automatic stopping criterion for the
Monte Carlo simulation. As soon as $m$ starts to decrease faster
in the course of the simulation, the routine makes more steps to ensure
equilibration, see Fig. \ref{Fig_MCS_vs_T_RF_xy}. This leads to the
observed jump to the demagnetized state. If a fixed number of MCS
is done at each temperature, the dependence $m(T)$ is smoother. The
temperature dependence of the system's energy $U(T)$ does not show
any singularity at the unblocking point -- there is only a region
with a larger slope that leads to a broad maximum of the heat capacity.
Such a behavior is similar to that of a system of noninteracting magnetic
particles with random directions of the uniaxial anisotropy axes.
Although there is exchange interaction between IM domains, it is weak
because for small $D_{R}/J$ IM domains are large and the exchange
connectivity in 2D is insufficient to produce many-body effects such
as spin-glass freezing.

\begin{figure}
\begin{centering}
\includegraphics[width=8cm]{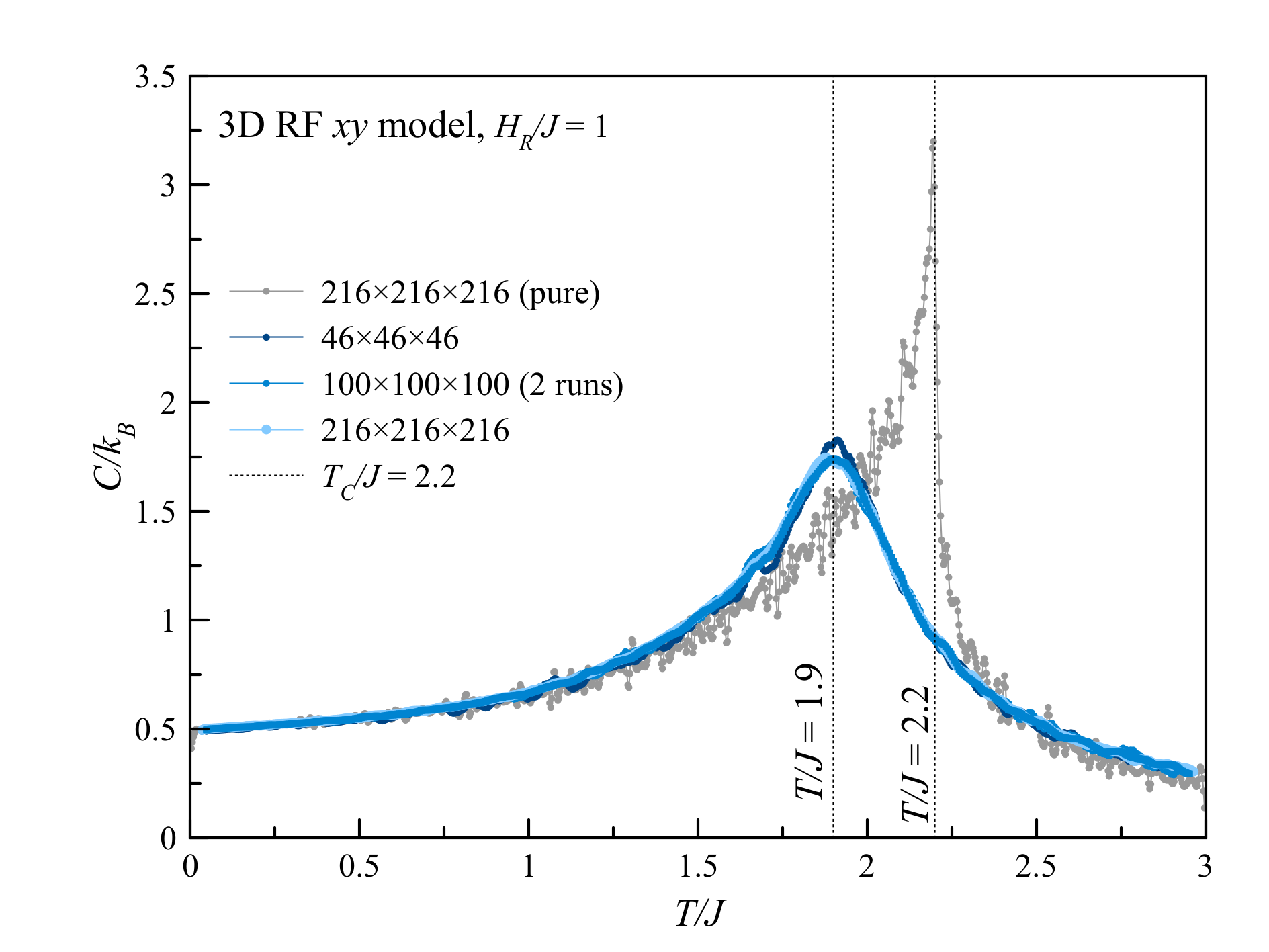}
\par\end{centering}
\caption{Temperature dependence of the heat capacity of the 3D $xy$ RF model,
compared with that of the pure system.  }\label{Fig_C_vs_T_pbc_3D_xy}

\end{figure}

The temperature dependence of the magnetization in 3D RA and RF models
was computed on lowering the temperature to check whether these systems
order spontaneously. Figure \ref{Fig_m_vs_T_3D} shows that the 3D
RA systems order on lowering $T$, while 3D RF models order only as
a finite-size effect, and ordering disappears in the limit of large
systems. Also for the RA model, the magnetization at low $T$ is decreasing
with increasing system size. However, the magnetization decrease is
small. Also, one should keep in mind that equilibrating a very large
system, especially in the presence of energy barriers, is difficult
and may take an exorbitant computing time. In real systems, there
are many kinds of weak interactions that affect ordering at large
distances. For instance, ferromagnets in the presence of the dipole-dipole
interactions form domains that reduce the magnetization of the system.
However, one does not say these materials do not ferromagnetically
order. To judge about ordering, one also needs to look at the singularities
at the transition point. In Fig. \ref{Fig_m_vs_T_3D} one can see
that there is a clear singularity of $m(T)$ in the RA model similar
to that for the pure system, and the phase transition temperature
$T_{C}/J=1.44$ is not changed by the moderate RA such as $D_{R}/J=1$.
To the contrary, the transition in the RF model is rounded, resembling
the pure ferromagnet in the applied field. Figure \ref{Fig_m_vs_T_100x100x100_DR_var}
shows that in the RA model, ordering occurs even for the strong RA
such as $D_{R}/J=3$ and 10. Here, one cannot say that the magnetization
at low temperatures might be a finite-size effect because for the
strong RA, the magnetic correlation radius $R_{f}$ is not large.

Absence and presence of the phase transition on temperature can also
be seen in the comparison plot of the energies of the RA and RF models
in Fig. \ref{Fig_U_vs_T}. The function $U(T)$ for the RA model has
the form usual for second-order phase transitions with a suppression
just below $T_{C}$. On the other hand, for the RF model, $U(T)$
has no pronounced features at any temperature. The heat capacity of
the 3D RA and RF models obtained by differentiating $U(T)$ is shown
in Fig. \ref{Fig_C_vs_T}. For the moderate-strength RA, $D_{R}/J=1$,
one has $C(T)$ diverging at the transition point, similarly to the
pure system. However, for the strong RA, $D_{R}/J=10$, the heat capacity
has a cusp typical to the spin-glass transition. In this case, the
system partially orders, see Fig. \ref{Fig_m_vs_T_100x100x100_DR_var},
so one can say is has combined properties of a ferromagnet and spin
glass. The RF model with $H_{R}/J=1$ also shows a maximum of $C(T)$
at $T/J=1.25$ in the form of a rounded cusp.

Finally, temperature dependences of physical quantities were also
computed for the 3D RA and RF two-component ($xy$) spin model. The
results are qualitatively similar to those of the Heisenberg model.
The temperature dependences of the magnetization on lowering temperature
in Fig. \ref{Fig_m_vs_T_3D_xy} are similar to those in Figs. \ref{Fig_m_vs_T_3D}
and \ref{Fig_m_vs_T_100x100x100_DR_var}. Note the large scatter of
the data at low $T$ for the RF $xy$ model, caused by different realizations
of the RF. Figure \ref{Fig_C_vs_T_pbc_3D_xy} shows the heat capacity
$C(T)$ of the 3D RF $xy$ model compared to that of the pure system.
Whereas in the pure system $C(T)$ diverges at the Curie point $T_{C}/J=2.2$,
the system with $H_{R}/J=1$ shows a cusp at an apparent spin-glass
freezing point $T_{F}/J=1.9$.

\section{Conclusion}

Extensive numerical experiments, including the energy minimization
at $T=0$ and Monte Carlo at $T>0$ on random-field and random-anisotropy
systems in 2D and 3D with 2 and 3 spin components were performed.
The results show that the lowest-energy states of these systems have
a significant magnetization, which contradicts the statements made
in the literature about the completely disordered ground states. In
RF systems, local energy levels are widely distributed, and higher-energy
states with lower magnetization contain singularities that are responsible
for the energy increase. On the other hand, a magnetized ground state
does not automatically lead to ordering with lowering temperature
because of pinned singularities.

The absence of ordering is expected in 2D systems having global continuous
symmetry. Moreover, it was shown here that the 2D RA system behaves
similarly to the assembly of non-interacting magnetic particles with
randomly directed uniaxial anisotropy axes. This is in agreement with
Ref. \citep{garchu22jpcm}, where blocking was observed instead of
a spin-glass freezing.

In 3D RF systems, ordering is prevented by the energy barriers and
pinned singularities. Instead of magnetic ordering, these systems
freeze into the correlated spin-glass state. The heat capacity of
these systems has a characteristic form with a cusp at the freezing
temperature.

On the other hand, the 3D RA model with a moderate-strength RA, $D_{R}/J\leq1$,
has local energy levels localized in the region with a high magnetization,
and these energy levels result from the energy minimization from any
initial conditions. A strong preference for the magnetized states
can be explained by the anisotropy of a spin having two opposite lowest-energy
states that makes the ordering of spins within a hemisphere preferable
in the case of the strong RA (for a weaker RA, the tendency to ordering
should be even stronger). In addition, for the moderate-to-weak RA,
singularities do not survive at $T=0$ because the exchange suppresses
them. As a result, 3D RA systems order on lowering the temperature.
For weak and moderate RA, relative to the exchange, the magnetic transition
is similar to that in the pure system, although a thorough investigation
of the critical behavior is needed. For a strong RA, the systems still
order, but the magnetization is substantially reduced, and there is
a considerable glassy component in this partially ordered state. The
critical behavior in this case is different, having the heat capacity
with a cusp instead of a divergence at the transition temperature.

In real materials, the anisotropy at the atomic scale is weak in comparison
to the exchange. However, materials composed of grains with the same
direction of RA within the grain have a much stronger effective RA.
The computations performed above for the atomic-scale RA can be done
for these materials as well.

\section*{Acknowledgment }

This work has been supported by the PSC-CUNY grant TRADA-54-119.

\bibliographystyle{aapmrev4-2}
\bibliography{C:/Users/einsc/Documents/BIBLIOTHEK/gar-static_randomness,C:/Users/einsc/Documents/BIBLIOTHEK/gar-own}

%aapmrev4-2.bst 2019-01-14 (MD) hand-edited version of aapmrev4-1.bst
\begin{thebibliography}{56}%
\makeatletter
\providecommand \@ifxundefined [1]{%
 \@ifx{#1\undefined}
}%
\providecommand \@ifnum [1]{%
 \ifnum #1\expandafter \@firstoftwo
 \else \expandafter \@secondoftwo
 \fi
}%
\providecommand \@ifx [1]{%
 \ifx #1\expandafter \@firstoftwo
 \else \expandafter \@secondoftwo
 \fi
}%
\providecommand \natexlab [1]{#1}%
\providecommand \enquote  [1]{``#1''}%
\providecommand \bibnamefont  [1]{#1}%
\providecommand \bibfnamefont [1]{#1}%
\providecommand \citenamefont [1]{#1}%
\providecommand \href@noop [0]{\@secondoftwo}%
\providecommand \href [0]{\begingroup \@sanitize@url \@href}%
\providecommand \@href[1]{\@@startlink{#1}\@@href}%
\providecommand \@@href[1]{\endgroup#1\@@endlink}%
\providecommand \@sanitize@url [0]{\catcode `\\12\catcode `\$12\catcode
  `\&12\catcode `\#12\catcode `\^12\catcode `\_12\catcode `\%12\relax}%
\providecommand \@@startlink[1]{}%
\providecommand \@@endlink[0]{}%
\providecommand \url  [0]{\begingroup\@sanitize@url \@url }%
\providecommand \@url [1]{\endgroup\@href {#1}{\urlprefix }}%
\providecommand \urlprefix  [0]{URL }%
\providecommand \Eprint [0]{\href }%
\providecommand \doibase [0]{https://doi.org/}%
\providecommand \selectlanguage [0]{\@gobble}%
\providecommand \bibinfo  [0]{\@secondoftwo}%
\providecommand \bibfield  [0]{\@secondoftwo}%
\providecommand \translation [1]{[#1]}%
\providecommand \BibitemOpen [0]{}%
\providecommand \bibitemStop [0]{}%
\providecommand \bibitemNoStop [0]{.\EOS\space}%
\providecommand \EOS [0]{\spacefactor3000\relax}%
\providecommand \BibitemShut  [1]{\csname bibitem#1\endcsname}%
\let\auto@bib@innerbib\@empty
%</preamble>
\bibitem [{\citenamefont {Mar\'{i}n}\ and\ \citenamefont
  {Hernando}(2000)}]{marher00jmmm}%
  \BibitemOpen
  \bibfield  {author} {\bibinfo {author} {\bibfnamefont {P.}~\bibnamefont
  {Mar\'{i}n}}\ and\ \bibinfo {author} {\bibfnamefont {A.}~\bibnamefont
  {Hernando}},\ }\href
  {https://doi.org/https://doi.org/10.1016/S0304-8853(00)00272-9} {\bibfield
  {journal} {\bibinfo  {journal} {Journal of Magnetism and Magnetic Materials}\
  }\textbf {\bibinfo {volume} {215-216}},\ \bibinfo {pages} {729} (\bibinfo
  {year} {2000})}\BibitemShut {NoStop}%
\bibitem [{\citenamefont {Billoni}, \citenamefont {Cannas},\ and\ \citenamefont
  {Tamarit}(2005)}]{bilcantam05prb}%
  \BibitemOpen
  \bibfield  {author} {\bibinfo {author} {\bibfnamefont {O.~V.}\ \bibnamefont
  {Billoni}}, \bibinfo {author} {\bibfnamefont {S.~A.}\ \bibnamefont
  {Cannas}},\ and\ \bibinfo {author} {\bibfnamefont {F.~A.}\ \bibnamefont
  {Tamarit}},\ }\href {https://doi.org/10.1103/PhysRevB.72.104407} {\bibfield
  {journal} {\bibinfo  {journal} {Phys. Rev. B}\ }\textbf {\bibinfo {volume}
  {72}},\ \bibinfo {pages} {104407} (\bibinfo {year} {2005})}\BibitemShut
  {NoStop}%
\bibitem [{\citenamefont {Larkin}(1970)}]{lar70jetp}%
  \BibitemOpen
  \bibfield  {author} {\bibinfo {author} {\bibfnamefont {A.~I.}\ \bibnamefont
  {Larkin}},\ }\href@noop {} {\bibfield  {journal} {\bibinfo  {journal} {Sov.
  Phys. JETP}\ }\textbf {\bibinfo {volume} {31}},\ \bibinfo {pages} {784}
  (\bibinfo {year} {1970})}\BibitemShut {NoStop}%
\bibitem [{\citenamefont {Chudnovsky}(1989)}]{chu89prb}%
  \BibitemOpen
  \bibfield  {author} {\bibinfo {author} {\bibfnamefont {E.~M.}\ \bibnamefont
  {Chudnovsky}},\ }\href {https://doi.org/10.1103/PhysRevB.40.11355} {\bibfield
   {journal} {\bibinfo  {journal} {Phys. Rev. B}\ }\textbf {\bibinfo {volume}
  {40}},\ \bibinfo {pages} {11355} (\bibinfo {year} {1989})}\BibitemShut
  {NoStop}%
\bibitem [{\citenamefont {Blatter}\ \emph {et~al.}(1994)\citenamefont
  {Blatter}, \citenamefont {Feigel'man}, \citenamefont {Geshkenbein},
  \citenamefont {Larkin},\ and\ \citenamefont
  {Vinokur}}]{blafeigeslarvin94rmp}%
  \BibitemOpen
  \bibfield  {author} {\bibinfo {author} {\bibfnamefont {G.}~\bibnamefont
  {Blatter}}, \bibinfo {author} {\bibfnamefont {M.~V.}\ \bibnamefont
  {Feigel'man}}, \bibinfo {author} {\bibfnamefont {V.~B.}\ \bibnamefont
  {Geshkenbein}}, \bibinfo {author} {\bibfnamefont {A.~I.}\ \bibnamefont
  {Larkin}},\ and\ \bibinfo {author} {\bibfnamefont {V.~M.}\ \bibnamefont
  {Vinokur}},\ }\href {https://doi.org/10.1103/RevModPhys.66.1125} {\bibfield
  {journal} {\bibinfo  {journal} {Rev. Mod. Phys.}\ }\textbf {\bibinfo {volume}
  {66}},\ \bibinfo {pages} {1125} (\bibinfo {year} {1994})}\BibitemShut
  {NoStop}%
\bibitem [{\citenamefont {Chudnovsky}(1986{\natexlab{a}})}]{chu86prb}%
  \BibitemOpen
  \bibfield  {author} {\bibinfo {author} {\bibfnamefont {E.~M.}\ \bibnamefont
  {Chudnovsky}},\ }\href {https://doi.org/10.1103/PhysRevB.33.245} {\bibfield
  {journal} {\bibinfo  {journal} {Phys. Rev. B}\ }\textbf {\bibinfo {volume}
  {33}},\ \bibinfo {pages} {245} (\bibinfo {year}
  {1986}{\natexlab{a}})}\BibitemShut {NoStop}%
\bibitem [{\citenamefont {Efetov}\ and\ \citenamefont
  {Larkin}(1977)}]{efelar77jetp}%
  \BibitemOpen
  \bibfield  {author} {\bibinfo {author} {\bibfnamefont {K.~B.}\ \bibnamefont
  {Efetov}}\ and\ \bibinfo {author} {\bibfnamefont {A.~I.}\ \bibnamefont
  {Larkin}},\ }\href@noop {} {\bibfield  {journal} {\bibinfo  {journal} {Sov.
  Phys. JETP}\ }\textbf {\bibinfo {volume} {45}},\ \bibinfo {pages} {1236}
  (\bibinfo {year} {1977})}\BibitemShut {NoStop}%
\bibitem [{\citenamefont {Volovik}(2008)}]{vol08jltp}%
  \BibitemOpen
  \bibfield  {author} {\bibinfo {author} {\bibfnamefont {G.}~\bibnamefont
  {Volovik}},\ }\href {https://doi.org/10.1007/s10909-007-9579-3} {\bibfield
  {journal} {\bibinfo  {journal} {J. Low. Temp. Phys.}\ }\textbf {\bibinfo
  {volume} {150}},\ \bibinfo {pages} {453} (\bibinfo {year}
  {2008})}\BibitemShut {NoStop}%
\bibitem [{\citenamefont {Li}\ \emph {et~al.}(2013)\citenamefont {Li},
  \citenamefont {Pollanen}, \citenamefont {Zimmerman}, \citenamefont {Collett},
  \citenamefont {Gannon},\ and\ \citenamefont
  {Halperin}}]{lipolzimcolganhal13nature}%
  \BibitemOpen
  \bibfield  {author} {\bibinfo {author} {\bibfnamefont {J.~I.}\ \bibnamefont
  {Li}}, \bibinfo {author} {\bibfnamefont {J.}~\bibnamefont {Pollanen}},
  \bibinfo {author} {\bibfnamefont {A.~M.}\ \bibnamefont {Zimmerman}}, \bibinfo
  {author} {\bibfnamefont {C.~A.}\ \bibnamefont {Collett}}, \bibinfo {author}
  {\bibfnamefont {W.~J.}\ \bibnamefont {Gannon}},\ and\ \bibinfo {author}
  {\bibfnamefont {W.~P.}\ \bibnamefont {Halperin}},\ }\href
  {https://doi.org/10.1038/nphys2806} {\bibfield  {journal} {\bibinfo
  {journal} {Nature Physics}\ }\textbf {\bibinfo {volume} {9}},\ \bibinfo
  {pages} {775} (\bibinfo {year} {2013})}\BibitemShut {NoStop}%
\bibitem [{\citenamefont {Seshadri}\ and\ \citenamefont
  {Westervelt}(1992{\natexlab{a}})}]{seswes92prb_1}%
  \BibitemOpen
  \bibfield  {author} {\bibinfo {author} {\bibfnamefont {R.}~\bibnamefont
  {Seshadri}}\ and\ \bibinfo {author} {\bibfnamefont {R.~M.}\ \bibnamefont
  {Westervelt}},\ }\href {https://doi.org/10.1103/PhysRevB.46.5142} {\bibfield
  {journal} {\bibinfo  {journal} {Phys. Rev. B}\ }\textbf {\bibinfo {volume}
  {46}},\ \bibinfo {pages} {5142} (\bibinfo {year}
  {1992}{\natexlab{a}})}\BibitemShut {NoStop}%
\bibitem [{\citenamefont {Seshadri}\ and\ \citenamefont
  {Westervelt}(1992{\natexlab{b}})}]{seswes92prb_2}%
  \BibitemOpen
  \bibfield  {author} {\bibinfo {author} {\bibfnamefont {R.}~\bibnamefont
  {Seshadri}}\ and\ \bibinfo {author} {\bibfnamefont {R.~M.}\ \bibnamefont
  {Westervelt}},\ }\href {https://doi.org/10.1103/PhysRevB.46.5150} {\bibfield
  {journal} {\bibinfo  {journal} {Phys. Rev. B}\ }\textbf {\bibinfo {volume}
  {46}},\ \bibinfo {pages} {5150} (\bibinfo {year}
  {1992}{\natexlab{b}})}\BibitemShut {NoStop}%
\bibitem [{\citenamefont {Bellini}\ \emph {et~al.}(1998)\citenamefont
  {Bellini}, \citenamefont {Clark}, \citenamefont {Degiorgio}, \citenamefont
  {Mantegazza},\ and\ \citenamefont {Natale}}]{belcladegmannat98pre}%
  \BibitemOpen
  \bibfield  {author} {\bibinfo {author} {\bibfnamefont {T.}~\bibnamefont
  {Bellini}}, \bibinfo {author} {\bibfnamefont {N.~A.}\ \bibnamefont {Clark}},
  \bibinfo {author} {\bibfnamefont {V.}~\bibnamefont {Degiorgio}}, \bibinfo
  {author} {\bibfnamefont {F.}~\bibnamefont {Mantegazza}},\ and\ \bibinfo
  {author} {\bibfnamefont {G.}~\bibnamefont {Natale}},\ }\href
  {https://doi.org/10.1103/PhysRevE.57.2996} {\bibfield  {journal} {\bibinfo
  {journal} {Phys. Rev. E}\ }\textbf {\bibinfo {volume} {57}},\ \bibinfo
  {pages} {2996} (\bibinfo {year} {1998})}\BibitemShut {NoStop}%
\bibitem [{\citenamefont {Harris}, \citenamefont {Plischke},\ and\
  \citenamefont {Zuckermann}(1973)}]{harplizuc73prl}%
  \BibitemOpen
  \bibfield  {author} {\bibinfo {author} {\bibfnamefont {R.}~\bibnamefont
  {Harris}}, \bibinfo {author} {\bibfnamefont {M.}~\bibnamefont {Plischke}},\
  and\ \bibinfo {author} {\bibfnamefont {M.~J.}\ \bibnamefont {Zuckermann}},\
  }\href {https://doi.org/10.1103/PhysRevLett.31.160} {\bibfield  {journal}
  {\bibinfo  {journal} {Phys. Rev. Lett.}\ }\textbf {\bibinfo {volume} {31}},\
  \bibinfo {pages} {160} (\bibinfo {year} {1973})}\BibitemShut {NoStop}%
\bibitem [{\citenamefont {Fishman}\ and\ \citenamefont
  {Aharony}(1979)}]{fisaha79jpcm}%
  \BibitemOpen
  \bibfield  {author} {\bibinfo {author} {\bibfnamefont {S.}~\bibnamefont
  {Fishman}}\ and\ \bibinfo {author} {\bibfnamefont {A.}~\bibnamefont
  {Aharony}},\ }\href {https://doi.org/10.1088/0022-3719/12/18/006} {\bibfield
  {journal} {\bibinfo  {journal} {Journal of Physics C: Solid State Physics}\
  }\textbf {\bibinfo {volume} {12}},\ \bibinfo {pages} {L729} (\bibinfo {year}
  {1979})}\BibitemShut {NoStop}%
\bibitem [{\citenamefont {Imry}\ and\ \citenamefont {Ma}(1975)}]{imrma75prl}%
  \BibitemOpen
  \bibfield  {author} {\bibinfo {author} {\bibfnamefont {Y.}~\bibnamefont
  {Imry}}\ and\ \bibinfo {author} {\bibfnamefont {S.-k.}\ \bibnamefont {Ma}},\
  }\href {https://doi.org/10.1103/PhysRevLett.35.1399} {\bibfield  {journal}
  {\bibinfo  {journal} {Phys. Rev. Lett.}\ }\textbf {\bibinfo {volume} {35}},\
  \bibinfo {pages} {1399} (\bibinfo {year} {1975})}\BibitemShut {NoStop}%
\bibitem [{\citenamefont {Aizenman}\ and\ \citenamefont
  {Wehr}(1989)}]{aizweh89prl}%
  \BibitemOpen
  \bibfield  {author} {\bibinfo {author} {\bibfnamefont {M.}~\bibnamefont
  {Aizenman}}\ and\ \bibinfo {author} {\bibfnamefont {J.}~\bibnamefont
  {Wehr}},\ }\href {https://doi.org/10.1103/PhysRevLett.62.2503} {\bibfield
  {journal} {\bibinfo  {journal} {Phys. Rev. Lett.}\ }\textbf {\bibinfo
  {volume} {62}},\ \bibinfo {pages} {2503} (\bibinfo {year}
  {1989})}\BibitemShut {NoStop}%
\bibitem [{\citenamefont {Aizenman}\ and\ \citenamefont
  {Wehr}(1990)}]{aizweh90cmp}%
  \BibitemOpen
  \bibfield  {author} {\bibinfo {author} {\bibfnamefont {M.}~\bibnamefont
  {Aizenman}}\ and\ \bibinfo {author} {\bibfnamefont {J.}~\bibnamefont
  {Wehr}},\ }\href {https://doi.org/10.1007/BF02096933} {\bibfield  {journal}
  {\bibinfo  {journal} {Commun.Math. Phys.}\ }\textbf {\bibinfo {volume}
  {130}},\ \bibinfo {pages} {489} (\bibinfo {year} {1990})}\BibitemShut
  {NoStop}%
\bibitem [{\citenamefont {Callen}, \citenamefont {Liu},\ and\ \citenamefont
  {Cullen}(1977)}]{calliucul77prb}%
  \BibitemOpen
  \bibfield  {author} {\bibinfo {author} {\bibfnamefont {E.}~\bibnamefont
  {Callen}}, \bibinfo {author} {\bibfnamefont {Y.~J.}\ \bibnamefont {Liu}},\
  and\ \bibinfo {author} {\bibfnamefont {J.~R.}\ \bibnamefont {Cullen}},\
  }\href {https://doi.org/10.1103/PhysRevB.16.263} {\bibfield  {journal}
  {\bibinfo  {journal} {Phys. Rev. B}\ }\textbf {\bibinfo {volume} {16}},\
  \bibinfo {pages} {263} (\bibinfo {year} {1977})}\BibitemShut {NoStop}%
\bibitem [{\citenamefont {Jayaprakash}\ and\ \citenamefont
  {Kirkpatrick}(1980)}]{jaykir80prb}%
  \BibitemOpen
  \bibfield  {author} {\bibinfo {author} {\bibfnamefont {C.}~\bibnamefont
  {Jayaprakash}}\ and\ \bibinfo {author} {\bibfnamefont {S.}~\bibnamefont
  {Kirkpatrick}},\ }\href {https://doi.org/10.1103/PhysRevB.21.4072} {\bibfield
   {journal} {\bibinfo  {journal} {Phys. Rev. B}\ }\textbf {\bibinfo {volume}
  {21}},\ \bibinfo {pages} {4072} (\bibinfo {year} {1980})}\BibitemShut
  {NoStop}%
\bibitem [{\citenamefont {Aharony}\ and\ \citenamefont
  {Pytte}(1980)}]{ahapyt80}%
  \BibitemOpen
  \bibfield  {author} {\bibinfo {author} {\bibfnamefont {A.}~\bibnamefont
  {Aharony}}\ and\ \bibinfo {author} {\bibfnamefont {E.}~\bibnamefont
  {Pytte}},\ }\href {https://doi.org/10.1103/PhysRevLett.45.1583} {\bibfield
  {journal} {\bibinfo  {journal} {Phys. Rev. Lett.}\ }\textbf {\bibinfo
  {volume} {45}},\ \bibinfo {pages} {1583} (\bibinfo {year}
  {1980})}\BibitemShut {NoStop}%
\bibitem [{\citenamefont {Aharony}\ and\ \citenamefont
  {Pytte}(1983)}]{ahapyt83}%
  \BibitemOpen
  \bibfield  {author} {\bibinfo {author} {\bibfnamefont {A.}~\bibnamefont
  {Aharony}}\ and\ \bibinfo {author} {\bibfnamefont {E.}~\bibnamefont
  {Pytte}},\ }\href {https://doi.org/10.1103/PhysRevB.27.5872} {\bibfield
  {journal} {\bibinfo  {journal} {Phys. Rev. B}\ }\textbf {\bibinfo {volume}
  {27}},\ \bibinfo {pages} {5872} (\bibinfo {year} {1983})}\BibitemShut
  {NoStop}%
\bibitem [{\citenamefont {Aharony}(1983)}]{aha83jmmm}%
  \BibitemOpen
  \bibfield  {author} {\bibinfo {author} {\bibfnamefont {A.}~\bibnamefont
  {Aharony}},\ }\href
  {https://doi.org/https://doi.org/10.1016/0304-8853(83)90959-9} {\bibfield
  {journal} {\bibinfo  {journal} {Journal of Magnetism and Magnetic Materials}\
  }\textbf {\bibinfo {volume} {31-34}},\ \bibinfo {pages} {1432} (\bibinfo
  {year} {1983})}\BibitemShut {NoStop}%
\bibitem [{\citenamefont {Dudka}, \citenamefont {Folk},\ and\ \citenamefont
  {Holovatch}(2005)}]{dudfolhol05jmmm}%
  \BibitemOpen
  \bibfield  {author} {\bibinfo {author} {\bibfnamefont {M.}~\bibnamefont
  {Dudka}}, \bibinfo {author} {\bibfnamefont {R.}~\bibnamefont {Folk}},\ and\
  \bibinfo {author} {\bibfnamefont {Y.}~\bibnamefont {Holovatch}},\ }\href
  {https://doi.org/https://doi.org/10.1016/j.jmmm.2004.12.028} {\bibfield
  {journal} {\bibinfo  {journal} {Journal of Magnetism and Magnetic Materials}\
  }\textbf {\bibinfo {volume} {294}},\ \bibinfo {pages} {305} (\bibinfo {year}
  {2005})}\BibitemShut {NoStop}%
\bibitem [{\citenamefont {Garanin}, \citenamefont {Chudnovsky},\ and\
  \citenamefont {Proctor}(2013)}]{garchupro13prb}%
  \BibitemOpen
  \bibfield  {author} {\bibinfo {author} {\bibfnamefont {D.~A.}\ \bibnamefont
  {Garanin}}, \bibinfo {author} {\bibfnamefont {E.~M.}\ \bibnamefont
  {Chudnovsky}},\ and\ \bibinfo {author} {\bibfnamefont {T.~C.}\ \bibnamefont
  {Proctor}},\ }\href {https://doi.org/10.1103/PhysRevB.88.224418} {\bibfield
  {journal} {\bibinfo  {journal} {Phys. Rev. B}\ }\textbf {\bibinfo {volume}
  {88}},\ \bibinfo {pages} {224418} (\bibinfo {year} {2013})}\BibitemShut
  {NoStop}%
\bibitem [{\citenamefont {Proctor}, \citenamefont {Garanin},\ and\
  \citenamefont {Chudnovsky}(2014)}]{progarchu14prl}%
  \BibitemOpen
  \bibfield  {author} {\bibinfo {author} {\bibfnamefont {T.~C.}\ \bibnamefont
  {Proctor}}, \bibinfo {author} {\bibfnamefont {D.~A.}\ \bibnamefont
  {Garanin}},\ and\ \bibinfo {author} {\bibfnamefont {E.~M.}\ \bibnamefont
  {Chudnovsky}},\ }\href {https://doi.org/10.1103/PhysRevLett.112.097201}
  {\bibfield  {journal} {\bibinfo  {journal} {Phys. Rev. Lett.}\ }\textbf
  {\bibinfo {volume} {112}},\ \bibinfo {pages} {097201} (\bibinfo {year}
  {2014})}\BibitemShut {NoStop}%
\bibitem [{\citenamefont {Garanin}\ and\ \citenamefont
  {Chudnovsky}(2015)}]{garchu15epjb}%
  \BibitemOpen
  \bibfield  {author} {\bibinfo {author} {\bibfnamefont {D.~A.}\ \bibnamefont
  {Garanin}}\ and\ \bibinfo {author} {\bibfnamefont {E.~M.}\ \bibnamefont
  {Chudnovsky}},\ }\href {https://doi.org/10.1140/epjb/e2015-50604-x}
  {\bibfield  {journal} {\bibinfo  {journal} {Eur. Phys. J. B}\ }\textbf
  {\bibinfo {volume} {88}},\ \bibinfo {pages} {81} (\bibinfo {year}
  {2015})}\BibitemShut {NoStop}%
\bibitem [{\citenamefont {Cardy}\ and\ \citenamefont
  {Ostlund}(1982)}]{carost82prb}%
  \BibitemOpen
  \bibfield  {author} {\bibinfo {author} {\bibfnamefont {J.~L.}\ \bibnamefont
  {Cardy}}\ and\ \bibinfo {author} {\bibfnamefont {S.}~\bibnamefont
  {Ostlund}},\ }\href {https://doi.org/10.1103/PhysRevB.25.6899} {\bibfield
  {journal} {\bibinfo  {journal} {Phys. Rev. B}\ }\textbf {\bibinfo {volume}
  {25}},\ \bibinfo {pages} {6899} (\bibinfo {year} {1982})}\BibitemShut
  {NoStop}%
\bibitem [{\citenamefont {Villain}\ and\ \citenamefont
  {Fernandez}(1984)}]{vilfer84zfb}%
  \BibitemOpen
  \bibfield  {author} {\bibinfo {author} {\bibfnamefont {J.}~\bibnamefont
  {Villain}}\ and\ \bibinfo {author} {\bibfnamefont {J.}~\bibnamefont
  {Fernandez}},\ }\href {https://doi.org/10.1007/BF01388065} {\bibfield
  {journal} {\bibinfo  {journal} {Z. Physik B}\ }\textbf {\bibinfo {volume}
  {54}},\ \bibinfo {pages} {139} (\bibinfo {year} {1984})}\BibitemShut
  {NoStop}%
\bibitem [{\citenamefont {Korshunov}(1993)}]{kor93prb}%
  \BibitemOpen
  \bibfield  {author} {\bibinfo {author} {\bibfnamefont {S.~E.}\ \bibnamefont
  {Korshunov}},\ }\href {https://doi.org/10.1103/PhysRevB.48.3969} {\bibfield
  {journal} {\bibinfo  {journal} {Phys. Rev. B}\ }\textbf {\bibinfo {volume}
  {48}},\ \bibinfo {pages} {3969} (\bibinfo {year} {1993})}\BibitemShut
  {NoStop}%
\bibitem [{\citenamefont {Giamarchi}\ and\ \citenamefont
  {Le~Doussal}(1994)}]{giadou94prl}%
  \BibitemOpen
  \bibfield  {author} {\bibinfo {author} {\bibfnamefont {T.}~\bibnamefont
  {Giamarchi}}\ and\ \bibinfo {author} {\bibfnamefont {P.}~\bibnamefont
  {Le~Doussal}},\ }\href {https://doi.org/10.1103/PhysRevLett.72.1530}
  {\bibfield  {journal} {\bibinfo  {journal} {Phys. Rev. Lett.}\ }\textbf
  {\bibinfo {volume} {72}},\ \bibinfo {pages} {1530} (\bibinfo {year}
  {1994})}\BibitemShut {NoStop}%
\bibitem [{\citenamefont {Giamarchi}\ and\ \citenamefont
  {Le~Doussal}(1995)}]{giadou95prb}%
  \BibitemOpen
  \bibfield  {author} {\bibinfo {author} {\bibfnamefont {T.}~\bibnamefont
  {Giamarchi}}\ and\ \bibinfo {author} {\bibfnamefont {P.}~\bibnamefont
  {Le~Doussal}},\ }\href {https://doi.org/10.1103/PhysRevB.52.1242} {\bibfield
  {journal} {\bibinfo  {journal} {Phys. Rev. B}\ }\textbf {\bibinfo {volume}
  {52}},\ \bibinfo {pages} {1242} (\bibinfo {year} {1995})}\BibitemShut
  {NoStop}%
\bibitem [{\citenamefont {Feldman}(2000)}]{fel00prb}%
  \BibitemOpen
  \bibfield  {author} {\bibinfo {author} {\bibfnamefont {D.~E.}\ \bibnamefont
  {Feldman}},\ }\href {https://doi.org/10.1103/PhysRevB.61.382} {\bibfield
  {journal} {\bibinfo  {journal} {Phys. Rev. B}\ }\textbf {\bibinfo {volume}
  {61}},\ \bibinfo {pages} {382} (\bibinfo {year} {2000})}\BibitemShut
  {NoStop}%
\bibitem [{\citenamefont {Feldman}(2001)}]{fel01ijmp}%
  \BibitemOpen
  \bibfield  {author} {\bibinfo {author} {\bibfnamefont {D.~E.}\ \bibnamefont
  {Feldman}},\ }\href {https://doi.org/10.1142/S0217979201006641} {\bibfield
  {journal} {\bibinfo  {journal} {International Journal of Modern Physics B}\
  }\textbf {\bibinfo {volume} {15}},\ \bibinfo {pages} {2945} (\bibinfo {year}
  {2001})}\BibitemShut {NoStop}%
\bibitem [{\citenamefont {Le~Doussal}(2006)}]{dou06prl}%
  \BibitemOpen
  \bibfield  {author} {\bibinfo {author} {\bibfnamefont {P.}~\bibnamefont
  {Le~Doussal}},\ }\href {https://doi.org/10.1103/PhysRevLett.96.235702}
  {\bibfield  {journal} {\bibinfo  {journal} {Phys. Rev. Lett.}\ }\textbf
  {\bibinfo {volume} {96}},\ \bibinfo {pages} {235702} (\bibinfo {year}
  {2006})}\BibitemShut {NoStop}%
\bibitem [{\citenamefont {Nattermann}\ and\ \citenamefont
  {Scheidl}(2000)}]{natsch00advphys}%
  \BibitemOpen
  \bibfield  {author} {\bibinfo {author} {\bibfnamefont {T.}~\bibnamefont
  {Nattermann}}\ and\ \bibinfo {author} {\bibfnamefont {S.}~\bibnamefont
  {Scheidl}},\ }\href {https://doi.org/10.1080/000187300412257} {\bibfield
  {journal} {\bibinfo  {journal} {Advances in Physics}\ }\textbf {\bibinfo
  {volume} {49}},\ \bibinfo {pages} {607} (\bibinfo {year} {2000})}\BibitemShut
  {NoStop}%
\bibitem [{\citenamefont {Garel}, \citenamefont {Iori},\ and\ \citenamefont
  {Orland}(1996)}]{gariororl96prb}%
  \BibitemOpen
  \bibfield  {author} {\bibinfo {author} {\bibfnamefont {T.}~\bibnamefont
  {Garel}}, \bibinfo {author} {\bibfnamefont {G.}~\bibnamefont {Iori}},\ and\
  \bibinfo {author} {\bibfnamefont {H.}~\bibnamefont {Orland}},\ }\href
  {https://doi.org/10.1103/PhysRevB.53.R2941} {\bibfield  {journal} {\bibinfo
  {journal} {Phys. Rev. B}\ }\textbf {\bibinfo {volume} {53}},\ \bibinfo
  {pages} {R2941} (\bibinfo {year} {1996})}\BibitemShut {NoStop}%
\bibitem [{\citenamefont {Gingras}\ and\ \citenamefont
  {Huse}(1996)}]{ginhus96prb}%
  \BibitemOpen
  \bibfield  {author} {\bibinfo {author} {\bibfnamefont {M.~J.~P.}\
  \bibnamefont {Gingras}}\ and\ \bibinfo {author} {\bibfnamefont {D.~A.}\
  \bibnamefont {Huse}},\ }\href {https://doi.org/10.1103/PhysRevB.53.15193}
  {\bibfield  {journal} {\bibinfo  {journal} {Phys. Rev. B}\ }\textbf {\bibinfo
  {volume} {53}},\ \bibinfo {pages} {15193} (\bibinfo {year}
  {1996})}\BibitemShut {NoStop}%
\bibitem [{\citenamefont {Fisch}(1998)}]{fis98prb}%
  \BibitemOpen
  \bibfield  {author} {\bibinfo {author} {\bibfnamefont {R.}~\bibnamefont
  {Fisch}},\ }\href {https://doi.org/10.1103/PhysRevB.58.5684} {\bibfield
  {journal} {\bibinfo  {journal} {Phys. Rev. B}\ }\textbf {\bibinfo {volume}
  {58}},\ \bibinfo {pages} {5684} (\bibinfo {year} {1998})}\BibitemShut
  {NoStop}%
\bibitem [{\citenamefont {Fisch}(2000)}]{fis00prb}%
  \BibitemOpen
  \bibfield  {author} {\bibinfo {author} {\bibfnamefont {R.}~\bibnamefont
  {Fisch}},\ }\href {https://doi.org/10.1103/PhysRevB.62.361} {\bibfield
  {journal} {\bibinfo  {journal} {Phys. Rev. B}\ }\textbf {\bibinfo {volume}
  {62}},\ \bibinfo {pages} {361} (\bibinfo {year} {2000})}\BibitemShut
  {NoStop}%
\bibitem [{\citenamefont {Itakura}(2003)}]{ita03prb}%
  \BibitemOpen
  \bibfield  {author} {\bibinfo {author} {\bibfnamefont {M.}~\bibnamefont
  {Itakura}},\ }\href {https://doi.org/10.1103/PhysRevB.68.100405} {\bibfield
  {journal} {\bibinfo  {journal} {Phys. Rev. B}\ }\textbf {\bibinfo {volume}
  {68}},\ \bibinfo {pages} {100405} (\bibinfo {year} {2003})}\BibitemShut
  {NoStop}%
\bibitem [{\citenamefont {Chudnovsky}\ and\ \citenamefont
  {Serota}(1982)}]{chuser82prb}%
  \BibitemOpen
  \bibfield  {author} {\bibinfo {author} {\bibfnamefont {E.~M.}\ \bibnamefont
  {Chudnovsky}}\ and\ \bibinfo {author} {\bibfnamefont {R.~A.}\ \bibnamefont
  {Serota}},\ }\href {https://doi.org/10.1103/PhysRevB.26.2697} {\bibfield
  {journal} {\bibinfo  {journal} {Phys. Rev. B}\ }\textbf {\bibinfo {volume}
  {26}},\ \bibinfo {pages} {2697} (\bibinfo {year} {1982})}\BibitemShut
  {NoStop}%
\bibitem [{\citenamefont {Chudnovsky}, \citenamefont {Saslow},\ and\
  \citenamefont {Serota}(1986)}]{chusasser86prb}%
  \BibitemOpen
  \bibfield  {author} {\bibinfo {author} {\bibfnamefont {E.~M.}\ \bibnamefont
  {Chudnovsky}}, \bibinfo {author} {\bibfnamefont {W.~M.}\ \bibnamefont
  {Saslow}},\ and\ \bibinfo {author} {\bibfnamefont {R.~A.}\ \bibnamefont
  {Serota}},\ }\href {https://doi.org/10.1103/PhysRevB.33.251} {\bibfield
  {journal} {\bibinfo  {journal} {Phys. Rev. B}\ }\textbf {\bibinfo {volume}
  {33}},\ \bibinfo {pages} {251} (\bibinfo {year} {1986})}\BibitemShut
  {NoStop}%
\bibitem [{\citenamefont {Chudnovsky}(1986{\natexlab{b}})}]{chu86prb_ra}%
  \BibitemOpen
  \bibfield  {author} {\bibinfo {author} {\bibfnamefont {E.~M.}\ \bibnamefont
  {Chudnovsky}},\ }\href {https://doi.org/10.1103/PhysRevB.33.2021} {\bibfield
  {journal} {\bibinfo  {journal} {Phys. Rev. B}\ }\textbf {\bibinfo {volume}
  {33}},\ \bibinfo {pages} {2021} (\bibinfo {year}
  {1986}{\natexlab{b}})}\BibitemShut {NoStop}%
\bibitem [{\citenamefont {Proctor}, \citenamefont {Chudnovsky},\ and\
  \citenamefont {Garanin}(2015)}]{prochugar15jmmm_RA}%
  \BibitemOpen
  \bibfield  {author} {\bibinfo {author} {\bibfnamefont {T.}~\bibnamefont
  {Proctor}}, \bibinfo {author} {\bibfnamefont {E.}~\bibnamefont
  {Chudnovsky}},\ and\ \bibinfo {author} {\bibfnamefont {D.}~\bibnamefont
  {Garanin}},\ }\href
  {https://doi.org/https://doi.org/10.1016/j.jmmm.2015.02.047} {\bibfield
  {journal} {\bibinfo  {journal} {Journal of Magnetism and Magnetic Materials}\
  }\textbf {\bibinfo {volume} {384}},\ \bibinfo {pages} {181} (\bibinfo {year}
  {2015})}\BibitemShut {NoStop}%
\bibitem [{\citenamefont {Garanin}\ and\ \citenamefont
  {Chudnovsky}(2022{\natexlab{a}})}]{garchu22jpcm}%
  \BibitemOpen
  \bibfield  {author} {\bibinfo {author} {\bibfnamefont {D.~A.}\ \bibnamefont
  {Garanin}}\ and\ \bibinfo {author} {\bibfnamefont {E.~M.}\ \bibnamefont
  {Chudnovsky}},\ }\href {https://doi.org/10.1088/1361-648x/ac684a} {\bibfield
  {journal} {\bibinfo  {journal} {Journal of Physics: Condensed Matter}\
  }\textbf {\bibinfo {volume} {34}},\ \bibinfo {pages} {285801} (\bibinfo
  {year} {2022}{\natexlab{a}})}\BibitemShut {NoStop}%
\bibitem [{\citenamefont {Garanin}\ and\ \citenamefont
  {Chudnovsky}(2021)}]{garchu21prb}%
  \BibitemOpen
  \bibfield  {author} {\bibinfo {author} {\bibfnamefont {D.~A.}\ \bibnamefont
  {Garanin}}\ and\ \bibinfo {author} {\bibfnamefont {E.~M.}\ \bibnamefont
  {Chudnovsky}},\ }\href {https://doi.org/10.1103/PhysRevB.103.214414}
  {\bibfield  {journal} {\bibinfo  {journal} {Phys. Rev. B}\ }\textbf {\bibinfo
  {volume} {103}},\ \bibinfo {pages} {214414} (\bibinfo {year}
  {2021})}\BibitemShut {NoStop}%
\bibitem [{\citenamefont {Garanin}\ and\ \citenamefont
  {Chudnovsky}(2022{\natexlab{b}})}]{garchu22prb}%
  \BibitemOpen
  \bibfield  {author} {\bibinfo {author} {\bibfnamefont {D.~A.}\ \bibnamefont
  {Garanin}}\ and\ \bibinfo {author} {\bibfnamefont {E.~M.}\ \bibnamefont
  {Chudnovsky}},\ }\href {https://doi.org/10.1103/PhysRevB.105.064402}
  {\bibfield  {journal} {\bibinfo  {journal} {Phys. Rev. B}\ }\textbf {\bibinfo
  {volume} {105}},\ \bibinfo {pages} {064402} (\bibinfo {year}
  {2022}{\natexlab{b}})}\BibitemShut {NoStop}%
\bibitem [{\citenamefont {Chudnovsky}\ and\ \citenamefont
  {Garanin}(2023)}]{garchu23prb_IP}%
  \BibitemOpen
  \bibfield  {author} {\bibinfo {author} {\bibfnamefont {E.~M.}\ \bibnamefont
  {Chudnovsky}}\ and\ \bibinfo {author} {\bibfnamefont {D.~A.}\ \bibnamefont
  {Garanin}},\ }\href {https://doi.org/10.1103/PhysRevB.107.224413} {\bibfield
  {journal} {\bibinfo  {journal} {Phys. Rev. B}\ }\textbf {\bibinfo {volume}
  {107}},\ \bibinfo {pages} {224413} (\bibinfo {year} {2023})}\BibitemShut
  {NoStop}%
\bibitem [{\citenamefont {Chudnovsky}\ and\ \citenamefont
  {Garanin}(2024)}]{chugar24prb_sc}%
  \BibitemOpen
  \bibfield  {author} {\bibinfo {author} {\bibfnamefont {E.~M.}\ \bibnamefont
  {Chudnovsky}}\ and\ \bibinfo {author} {\bibfnamefont {D.~A.}\ \bibnamefont
  {Garanin}},\ }\href {https://doi.org/10.1103/PhysRevB.109.054429} {\bibfield
  {journal} {\bibinfo  {journal} {Phys. Rev. B}\ }\textbf {\bibinfo {volume}
  {109}},\ \bibinfo {pages} {054429} (\bibinfo {year} {2024})}\BibitemShut
  {NoStop}%
\bibitem [{\citenamefont {Dickman}\ and\ \citenamefont
  {Chudnovsky}(1991)}]{dicchu91prb}%
  \BibitemOpen
  \bibfield  {author} {\bibinfo {author} {\bibfnamefont {R.}~\bibnamefont
  {Dickman}}\ and\ \bibinfo {author} {\bibfnamefont {E.~M.}\ \bibnamefont
  {Chudnovsky}},\ }\href {https://doi.org/10.1103/PhysRevB.44.4397} {\bibfield
  {journal} {\bibinfo  {journal} {Phys. Rev. B}\ }\textbf {\bibinfo {volume}
  {44}},\ \bibinfo {pages} {4397} (\bibinfo {year} {1991})}\BibitemShut
  {NoStop}%
\bibitem [{\citenamefont {Garanin}\ and\ \citenamefont
  {Chudnovsky}(2024)}]{garchu24epl_sc}%
  \BibitemOpen
  \bibfield  {author} {\bibinfo {author} {\bibfnamefont {D.~A.}\ \bibnamefont
  {Garanin}}\ and\ \bibinfo {author} {\bibfnamefont {E.~M.}\ \bibnamefont
  {Chudnovsky}},\ }\href {https://doi.org/10.1209/0295-5075/ad8372} {\bibfield
  {journal} {\bibinfo  {journal} {Europhysics Letters}\ }\textbf {\bibinfo
  {volume} {148}},\ \bibinfo {pages} {26001} (\bibinfo {year}
  {2024})}\BibitemShut {NoStop}%
\bibitem [{\citenamefont {Garanin}(2021)}]{gar21pre}%
  \BibitemOpen
  \bibfield  {author} {\bibinfo {author} {\bibfnamefont {D.~A.}\ \bibnamefont
  {Garanin}},\ }\href {https://doi.org/10.1103/PhysRevE.104.055306} {\bibfield
  {journal} {\bibinfo  {journal} {Phys. Rev. E}\ }\textbf {\bibinfo {volume}
  {104}},\ \bibinfo {pages} {055306} (\bibinfo {year} {2021})}\BibitemShut
  {NoStop}%
\bibitem [{\citenamefont {Chudnovsky}\ and\ \citenamefont
  {Garanin}(2018{\natexlab{a}})}]{chugar18prl}%
  \BibitemOpen
  \bibfield  {author} {\bibinfo {author} {\bibfnamefont {E.~M.}\ \bibnamefont
  {Chudnovsky}}\ and\ \bibinfo {author} {\bibfnamefont {D.~A.}\ \bibnamefont
  {Garanin}},\ }\href {https://doi.org/10.1103/PhysRevLett.121.017201}
  {\bibfield  {journal} {\bibinfo  {journal} {Phys. Rev. Lett.}\ }\textbf
  {\bibinfo {volume} {121}},\ \bibinfo {pages} {017201} (\bibinfo {year}
  {2018}{\natexlab{a}})}\BibitemShut {NoStop}%
\bibitem [{\citenamefont {Chudnovsky}\ and\ \citenamefont
  {Garanin}(2018{\natexlab{b}})}]{chugar18njp}%
  \BibitemOpen
  \bibfield  {author} {\bibinfo {author} {\bibfnamefont {E.~M.}\ \bibnamefont
  {Chudnovsky}}\ and\ \bibinfo {author} {\bibfnamefont {D.~A.}\ \bibnamefont
  {Garanin}},\ }\href {https://doi.org/10.1088/1367-2630/aab576} {\bibfield
  {journal} {\bibinfo  {journal} {New Journal of Physics}\ }\textbf {\bibinfo
  {volume} {20}},\ \bibinfo {pages} {033006} (\bibinfo {year}
  {2018}{\natexlab{b}})}\BibitemShut {NoStop}%
\bibitem [{\citenamefont {Döring}(1968)}]{doe68jap}%
  \BibitemOpen
  \bibfield  {author} {\bibinfo {author} {\bibfnamefont {W.}~\bibnamefont
  {Döring}},\ }\href {https://doi.org/10.1063/1.1656144} {\bibfield  {journal}
  {\bibinfo  {journal} {Journal of Applied Physics}\ }\textbf {\bibinfo
  {volume} {39}},\ \bibinfo {pages} {1006} (\bibinfo {year}
  {1968})}\BibitemShut {NoStop}%
\bibitem [{\citenamefont {Kuchkin}\ \emph {et~al.}(2025)\citenamefont
  {Kuchkin}, \citenamefont {Haller}, \citenamefont {Li$\check{\rm s}\check{\rm
  c}$\'{a}k}, \citenamefont {Adams}, \citenamefont {Rai}, \citenamefont
  {Sinaga}, \citenamefont {Michels},\ and\ \citenamefont
  {Schmidt}}]{kuchkinetal25prr}%
  \BibitemOpen
  \bibfield  {author} {\bibinfo {author} {\bibfnamefont {V.~M.}\ \bibnamefont
  {Kuchkin}}, \bibinfo {author} {\bibfnamefont {A.}~\bibnamefont {Haller}},
  \bibinfo {author} {\bibfnamefont {S.}~\bibnamefont {Li$\check{\rm
  s}\check{\rm c}$\'{a}k}}, \bibinfo {author} {\bibfnamefont {M.~P.}\
  \bibnamefont {Adams}}, \bibinfo {author} {\bibfnamefont {V.}~\bibnamefont
  {Rai}}, \bibinfo {author} {\bibfnamefont {E.~P.}\ \bibnamefont {Sinaga}},
  \bibinfo {author} {\bibfnamefont {A.}~\bibnamefont {Michels}},\ and\ \bibinfo
  {author} {\bibfnamefont {T.~L.}\ \bibnamefont {Schmidt}},\ }\href
  {https://doi.org/10.1103/PhysRevResearch.7.013195} {\bibfield  {journal}
  {\bibinfo  {journal} {Phys. Rev. Res.}\ }\textbf {\bibinfo {volume} {7}},\
  \bibinfo {pages} {013195} (\bibinfo {year} {2025})}\BibitemShut {NoStop}%
\end{thebibliography}%

\end{document}